\documentclass{aastex61}

\usepackage{amssymb,graphicx}

\begin{document}

\title{The Growth of the Density Fluctuations in the Scale-Invariant Vacuum Theory}

\correspondingauthor{Andre Maeder}
\email{andre.maeder@unige.ch}

\author[0000-0001-8744-0444]{Andre Maeder} 
\affiliation{Geneva Observatory \\
chemin des Maillettes, 51 \\
CH-1290 Sauverny, Switzerland}

\author[0000-0002-2022-6432]{Vesselin G. Gueorguiev}
\affiliation{Institute for Advanced Physical Studies, \\
Office 217, Building 2,\\
21 Montevideo Street, \\
Sofia 1618, Bulgaria}
\affiliation{ Ronin Institute for Independent Scholarship, \\
127 Haddon Pl.,\\
Montclair, NJ 07043, USA}

\begin{abstract}
The growth of the density fluctuations is considered to be an important cosmological test. 
In the standard model, for a matter dominated universe, 
the growth of the density perturbations evolves with redshift $z$ like  $(\frac{1}{1+z})^s$  with $s=1$. 
This is not fast enough to form galaxies and to account for the observed present-day inhomogeneities. 
This problem is usually resolved by assuming that at the recombination epoch the baryons settle down
in the potential well of the dark matter previously assembled during the radiation era of the universe.
This view is challenged in the present paper by using the recently proposed model of a scale-invariant 
framework for cosmology, the Scale-Invariant Vacuum Theory \cite{Maeder17a, Maeder17b, Maeder17c}, 
that enlarges the invariance group subtending the theory of the gravitation.

From the continuity equation, the Euler and Poisson equations are written in the scale-invariant framework,
the equation governing the growth of density fluctuations  $\delta$ is obtained. Starting from $\delta = 10^{-5}$
at a redshift around 1000, numerical solutions for various density background  are obtained.
The growth of density fluctuations is much faster than in the standard EdS model. 
The $s$ values are in the range from 2.7 to 3.9 for $\Omega_{\mathrm{m}}$ between 0.30 and 0.02. 
This enables the density fluctuations to enter the nonlinear regime with $\delta > 1$ 
long before the present time, typically at redshifts of about 10, 
without requiring the presence of dark matter.
\end{abstract}





\section{Introduction} \label{sec:intro}

The growth of density fluctuations in the early Universe is a fundamental  cosmological problem.
It has been  studied both by analytical developments, see for example  textbooks by
\citet{Coles02}, \citet{Durrer08} and \citet{Theuns16} and references therein and also by numerical simulations,
as mentioned below. Let us first briefly recall the basic points of the problem in the standard theory.
During the radiation dominated era, it is considered that 
 the sub-horizon baryonic 
density perturbations do not grow and only the perturbations 
smaller than the horizon distance (to be causally connected) and 
larger than a minimum mass (the Silk mass) are not rapidly erased by photon diffusion \citep{Silk68}.
At the time of recombination, at a redshift $z \approx 10^3$,
the observed temperature fluctuations of the CMB are of the order of $\delta T/T \approx 10^{-5}$.
 At this time, the fluctuations of the baryonic density $\delta \rho/\rho$
are of the same order, since up to recombination the matter and radiation were closely coupled. 
From this time onward, the growth of the density fluctuations $\delta \, \equiv \, \delta \rho/\rho$ 
in the linear regime is determined by the well-known equation \citep{Peebles80},
\begin{equation}
\ddot{\delta} + 2 \, H \, \dot{\delta} \,
= \, 4 \, \pi \, G \, \rho \,\delta \, ,
\label{std}
\end{equation}
\noindent 
where $H$ is the Hubble ``constant" at the time considered and  
 $\varrho$ the density. In the standard model for a dust Universe (when the pressure $p=0$),
$\delta$ goes like $t^{2/3} \, \propto \, a(t)\, \propto \, (1+z)^{-1}$, where $a(t)$ is the expansion factor 
and $z$ the redshift. Thus, starting from perturbations with an amplitude of
 about $10^{-5}$, a growth by a factor of $10^3$, since the time of recombination,
  would lead to fluctuations of order $10^{-2}$ at the present epoch. 
 This  is by several orders of magnitude smaller than the large non-linear structures presently observed in the Universe 
 \citep{Ostriker93}.
 
 The above discrepancy is one of the main reason, along with the flat rotation curves of spiral galaxies, for  the introduction 
  of dark matter, which was not more than a hypothesis   at the beginning of such studies.
   The arguments of the standard theory have thus become  the following ones.
 While baryonic perturbations do not amplify during the radiation era,
 the perturbations of the assumed dark matter (subject only to gravitational
 interaction) would be growing during the radiation era. Then, it is  considered  that after recombination
 the baryons are falling into the potential wells previously created by 
 the dark matter. Their growth towards the present highly contrasted structures are
  thus  strongly boosted. The cold dark matter (CDM) scenario and its related 
astronomical tests have been studied by \citet{Ostriker93}, while numerical simulations, techniques, and results
have been reviewed by \citet{Bertschinger98} and more recently by \citet{Sommerville15}.

In the present work, we study the growth of the density fluctuations in a scale-invariant  framework, in particular
for examining whether the hypothesis of dark matter is also necessary within this context. We also want to study the effect 
of the matter density $\Omega_{\mathrm{m}}$ and to estimate the redshift of galaxy formation.
A  presentation of the basis of the scale-invariant theory is given in Section \ref{base}.
In Section \ref{theory1}, the three  basic equations for the growth of density fluctuations,
the equation of continuity, the equations of Euler and Poisson are obtained in the scale-invariant context and then
applied to density perturbations  in Section  \ref{theory2}. 
In Section \ref{num}, numerical solutions of the equation for the growth of density fluctuations are obtained and discussed.
Section \ref{concl} gives the conclusions.\\

\section{Theoretical basis of the scale-invariant field equations}  \label{base}

\subsection{The integrable Weyl's geometry}

Weyl's geometry offers a consistent framework  for incorporating scale invariance in the equations of gravitation. It was proposed 
by Hermann Weyl  \citep{Weyl23} to interpret the electromagnetism  by an interesting generalization of the Riemann 
spaces  and was further clearly developed by \citet{Eddington23} and \citet{Dirac73}. It is well known that the Maxwell 
equations for electromagnetism are scale invariant in the empty space.
Weyl's space possesses a metrical determination with 
the quadratic form
\begin{equation}
ds^2 \, = \, g_{\mu \nu}(x) dx^\mu dx^\nu \, .
\label{ds}
\end{equation} 
\noindent
In addition, it allows the possibility  
that a vector of length  $\ell$ attached at a point  $P$  of coordinates
$x^\mu$ changes its length to $\ell +\delta \ell$  when transported 
by parallel displacement to a point $P'$ of coordinates $x^\mu+\delta x^\mu$.
The change in length $\delta \ell$ is proportional to $\ell$ and to $\delta x^\mu$,
\begin{equation}
\delta \ell \, = \, \ell \kappa_\mu \delta x^\mu \, ,
\label{k1}
\end{equation}
\noindent
where $\kappa_\mu$ is called the coefficient of metrical connection. 
In Weyl's geometry, $\kappa_\mu$  are fundamental coefficients as are the $g_{\mu \nu}$. 
It is also assumed that the lengths can undergo gauge change encoded by a multiplicative scale factor $\lambda(x)$: 
\begin{equation}
\ell' \, = \, \lambda(x) \, \ell \, . 
\label{l1}
\end{equation}
Let us examine how does  $\delta \ell'$  change. To the first order in $\delta x^\nu$, one has,
\begin{eqnarray}
\ell' +\delta \ell'\,=  \, (\ell+ \delta \ell) \lambda (x+\delta x) =
 (\ell+ \delta \ell) \lambda (x) + \ell \frac{\partial \lambda}{\partial x^\nu} \delta x^\nu \, , \\
 \delta \ell' \,= \, \lambda \delta \ell+ \ell \lambda_{, \nu} \delta x^\nu=
  \lambda \ell \kappa_\nu \delta x^\nu+  \ell \lambda_{, \nu} \delta x^\nu= 
  \lambda \ell \left( \kappa _\nu +\Phi_{, \nu} \right) \delta x^\nu \, ,
 \end{eqnarray}
 \noindent
 where the notation $ \lambda_{, \nu}= \frac{\partial \lambda}{\partial x^\nu}$ is used along with the expressions:
 \begin{equation}
 \Phi \, = \ \ln \lambda\, , \quad \quad \mathrm{and}  \quad \kappa'_\nu = 
 \kappa_\nu + \Phi _{, \nu}=\kappa_{\nu}+\partial_\nu \ln \lambda \, .
 \label{k2}
 \end{equation}
 \noindent
 If  the vector is  parallel transported along a closed loop, the total change of the length of the vector 
 can be written as:
 \begin{equation}
 \Delta \ell \, = \, \ell \left( \partial_\nu \kappa_{\mu}  - 
  \partial_\mu \kappa_{\mu} \right) \, \sigma^{\mu \nu} \, ,
 \label{boucle}
 \end{equation}
 \noindent
where $ \sigma^{\mu \nu}=dx^{\mu} \wedge dx^{\nu}$ 
is an infinitesimal surface element corresponding to the edges  $dx^{\mu}$ and $dx^{\nu}$.
In the above expression, the tensor  $F_{\mu \nu} \, = \,\left(\kappa_{\mu,\nu} -\kappa_{\nu,\mu}\right)$ was identified by Weyl
with the electromagnetic field.  The above form of the Weyl's geometry implies non-integrable length, which means that 
different atoms, due to their different world lines, would have different properties and thus will emit at different frequencies. 
This was the reason for Einstein's objection against Weyl's geometry, as recalled by \citet{Canu77}.
 
The above objection does not hold if one considers the so-called integrable Weyl's geometry, which forms a 
consistent framework for the study of gravitation as emphasized by \citet{Canu77} and \citet{BouvierM78}.
Let us consider that the framework of functions denoted by primes is a Riemann space as in GR. 
That is, $\kappa'_\nu=0$, and therefore:
 \begin{equation}
 \kappa_\nu \, = \, - \Phi_{, \nu} \, = \, - \frac{\partial \ln \lambda}{\partial x^\nu} \, .
 \label{k3}
 \end{equation}
 \noindent
The above condition means that the metrical connection $\kappa_\nu$ is the gradient of a scalar field 
$(\Phi=\ln\lambda)$, as seen above.
Thus, $\kappa_\nu dx^\nu$ is an exact differential and therefore:
 \begin{equation}
 \partial_\nu \kappa_{\mu}\, = \,  \partial_\mu \kappa_{\mu} \, ,
 \label{d2}
 \end{equation}
 \noindent
which according to Equation (\ref{boucle}) implies that the parallel displacement
of a vector along a closed loop does not change its length.\\
 
The interesting mathematical tools of Weyl's geometry  also works in the integrable form of this geometry. 
By considering the relation
\begin{equation}
ds' \, = \, \lambda(x^\mu) \, ds \, ,
\label{lambda}
\end{equation}
where  $ds'$ refers to the line element in the framework of the General Relativity (GR) 
while $ds$ to the more general Weyl's geometry framework.
Where $ds$ has the metrical form (\ref{ds}) and $ds'= g'_{\mu \nu} dx^\mu dx^\nu$. 
Thus, the integrable Weyl's space defined by $g_{\mu \nu}$ 
is conformally equivalent to a Riemann space (pseudo-space) defined by $g'_{\mu \nu}$
via the $\lambda$ mapping:
\begin{equation}
 g'_{\mu \nu} \, = \, \lambda^2 \, g_{\mu \nu} \, .
 \label{conformal}
 \end{equation}

\noindent
Scalars, vectors, or tensors  that  transform like
\begin{equation}
Y'^{\, \nu   }_{\mu}  \, =  \, \lambda^{n} \, Y^{\nu}_{\mu} \, ,
\label{co}
\end{equation}
\noindent
 are respectively called co-scalars, co-vectors, or  co-tensors  of power $n$.  
 If $n=0$, one has an in-scalar, in-vector or in-tensor, such objects are
 invariant upon a scale transformation  by a factor $\lambda$ like in (\ref{co}). Scale covariance  refers to a transformation 
 with  powers $n$ different from zero, while the term scale invariance is reserved for cases with  $n=0$.

The Cosmological Principle   demands that the scale factor   only  depends on time.
A co-tensor analysis has been developed by the above mentioned authors \citep{Weyl23,Eddington23,Dirac73, Canu77}.
\citet{BouvierM78}  have also derived the equation of geodesics from an action principle and shown its consistency
with the notion of shortest distance between two points in this context, they have also rediscussed the notion of parallel displacement,
of isometries and Killing vectors in the Integrable Weyl's geometry. 
In general, the derivative of a scale-invariant object is not scale invariant.
Thus,  scale covariant derivatives of the first and second order have been developed preserving  scale covariance.
For example, the co-covariant derivatives  $A_{\mu * \nu}$ and $A^{\mu}_{ * \nu}$ of a co-vector $A_\mu$ become
\begin{eqnarray}
A_{\mu * \nu} \, &=& \, \partial_\nu A_\mu - ^*\Gamma^{\alpha}_{\mu \nu} A_{\alpha} -n \kappa_\nu A_\mu  \, , \\ [2mm]
A^{\mu}_{ * \nu} \, &=& \, \partial_\nu A^\mu + ^*\Gamma^{\mu}_{\nu \alpha} A^{\alpha} -n \kappa_\nu A^\mu  \,  \\ [2mm]
\mathrm{with} \quad  ^*\Gamma^{\alpha}_{\mu \nu} &=& \Gamma^{\alpha}_{\mu \nu} 
+g_{\mu \nu}\kappa^\alpha \,- g^\alpha_\mu \kappa_\nu - g^\alpha_\nu \kappa_\mu .
\end{eqnarray}
\noindent
Here $^*\Gamma^{\alpha}_{\mu \nu}$ is a modified Christoffel symbol, 
while $\Gamma^{\alpha}_{\mu \nu}$ is the usual Christoffel symbol. 
The  Riemann curvature tensor $R^{\nu}_{\mu \lambda \rho}$,  
its contracted form, the Ricci tensor  $R^{\nu}_{\mu}$,  and the scalar
curvature $R$ also have their corresponding scale-covariant expressions:
\begin{equation}
R^{\nu}_{\mu} = R'^{\nu}_{\mu}  - \kappa^{; \nu}_{\mu}  - \kappa^{ \nu}_{;\mu}
 - g^{\nu}_{\mu}\kappa^{ \alpha}_{;\alpha}  -2 \kappa_{\mu} \kappa^{\nu}
+ 2 g^{\nu}_{\mu}\kappa^{ \alpha} \kappa_{ \alpha}  \, .
\label{RC}
\end{equation}
\begin{equation}
R \, = \, R' -6 \kappa^{\alpha}_{; \alpha}+6 \kappa^{\alpha} \kappa_{\alpha} \, .
\label{RRR}
\end{equation}
\noindent
Here the terms with a prime are the usual expressions in the Riemann geometry, 
and the semicolon ` ;  '  indicates the usual covariant derivative with respect to the relevant coordinate.
\citet{Canu77} provided a  short summary of the scale-covariant tensor analysis.
The main difference with the standard tensor analysis is that all these ``new''  expressions 
contain terms depending  on the coefficient of metrical connection $\kappa_{\nu}$  given by the above Equation (\ref{k3}).

\subsection{The general scale-invariant field equation and the gauge fixing}
The general field equation can be obtained in at least three different ways as shown by \citet{Canu77}. 
The first approach is in the spirit of the previous expressions, the first member of the field equation may be written.
The second member of the equation containing $T_{\mu \nu}$ must be scale invariant as is the first one. 
As shown below, this imposes some conditions on the scaling of pressures and densities.
The second way uses that the Equation (\ref{conformal}) represents a conformal transformation of the usual field equations, 
thus, the  corresponding transformations of the Ricci tensor and curvature scalar may be applied. 
The third way is based on the application of an action principle  \citep{Canu77} and also leads to the same field equation. 
The general field equation is usually written as
\begin{eqnarray}
R'_{\mu \nu}   -  \frac{1}{2}  \ g_{\mu \nu} R' - \kappa_{\mu ;\nu}  - \kappa_{ \nu ;\mu}
 -2 \kappa_{\mu} \kappa_ {\nu}   
+ 2 g_{\mu \nu} \kappa^{ \alpha}_{;\alpha}
 - g_{\mu \nu}\kappa^{ \alpha} \kappa_{ \alpha}   =
-8 \pi G T_{\mu \nu} - \lambda^2 \Lambda_{\mathrm{E}}  \, g_{\mu \nu} \, ,
\label{field}
\end{eqnarray}
\noindent
where $G$ is the gravitational constant (taken here as a true constant) and
 $\Lambda_{\mathrm{E}}$ the Einstein cosmological constant. 
The  scale-invariant energy-momentum tensor defined by 
$T_{\mu \nu}= T '_{\mu \nu}$ implies  \citep{Canu77}, 
\begin{equation}
 ( p+\varrho) u_{\mu} u_{\nu} -g_{\mu \nu } p =
   ( p'+\varrho') u'_{\mu} u'_{\nu} -g'_{\mu \nu } p' \, .
\end{equation}
The velocities $u'^{\mu}$ and $u'_{\mu}$ transform like
\begin{eqnarray}
u'^{\mu}=\frac{dx^{\mu}}{ds'}=\lambda^{-1}  \frac{dx^{\mu}}{ds}=  \lambda^{-1} u^{\mu} \, ,  \nonumber \\
  \; \mathrm{and} \; \;
u'_{\mu}=g'_{\mu \nu} u'^{\nu}=\lambda^2 g_{\mu \nu} \lambda^{-1} u^{\nu} = \lambda \, u_{\mu} \, .
\label{pl1}
\end{eqnarray}
\noindent
Thus, one has the following transformations for  the energy-momentum tensor and then for $p$ and $\rho$
\begin{equation}
( p+\varrho) u_{\mu} u_{\nu} -g_{\mu \nu } p  =
 ( p'+\varrho') \lambda^2 u_{\mu} u_{\nu} - \lambda^2 g_{\mu \nu } p' \, ,
 \label{pl}
\end{equation}
\begin{equation}
 p =  p'  \, \lambda^2    \, \quad \mathrm{and} \quad   \varrho =  \varrho'  \, \lambda^2 \, .
\label{ro2}
\end{equation}
\noindent
The pressure and density are therefore not scale-invariant but are so-called co-scalars of power  $n=-2$. 
The term containing the cosmological constant now writes  
$ -\lambda^2 \Lambda_{\mathrm{E}}  \, g_{\mu \nu}$, it is thus also a co-tensor of power $n=-2$.\\

It is not always realized that the field  equations of GR do have the property of scale invariance, as it is when
the cosmological constant $\Lambda$ is  equal to zero,  {\it{i.e.}} for a zero energy-density of the vacuum space, 
while for a non-zero $\Lambda$ the field equation is not scale-invariant. However, the above framework 
allows   scale invariance   together with a non-zero cosmological constant. 
In this context, it is worth to recall the interesting  remark by \citet{Bondi90}, who pointed out that 
 ``Einstein's disenchantment with the cosmological constant was partially motivated by a desire to preserve 
 scale-invariance of the empty space Einstein equations''.  The integrable Weyl geometry represents 
an interesting generalization of Riemann' geometry. As noted by \citet{Dirac73}, 
``it appears as one of the fundamental principles in Nature that the equations expressing basic laws 
should be invariant under the widest possible group of transformations''. Clearly, this is by no means a
proof of the validity of the scale-invariant vacuum theory and the comparison with the observations will decide whether it is true or not.
Such a comparison  is precisely the aim of the present work.

\subsection{Fixing the gauge within the Scale-Invariant Vacuum Theory}  \label{fix}

The field equation (\ref{field}) is undetermined and the same remark applies to the corresponding cosmological equations.
As a matter of fact, the same problem appears in General Relativity. As pointed out by \citet{Canu77}, 
the under-determinacy of GR is resolved by the choice of coordinates conditions. In the scale-invariant theory,
one needs to impose some gauging conditions to define the scale factor $\lambda$.
\citet{Dirac73} and \citet{Canu77} invoked the so-called Large Number Hypothesis to fix $\lambda$.
We do not follow this choice and consider a different physical condition.

The term  $ \lambda^2 \Lambda_{\mathrm{E}}$ represents the energy density
of the empty space in the scale-invariant context and we now make the specific hypothesis
that {\it{the properties of the empty space, at macroscopic scales, are scale invariant}} \citep{Maeder17a}.  
In this connection, we point out  that the usual equation of state for  the vacuum
$P_{\mathrm{vac}} = -\varrho_{\mathrm{vac}}$ is precisely what is permitting $\varrho_{\mathrm{vac}}$
to remain constant for an adiabatic expansion or contraction \citep{Carr92}.
At the quantum level, the scale invariance of the vacuum  does not necessarily apply, however in the same way
as one may use Einstein's theory at large-scales, even if it does not apply at the quantum level, 
we do consider that the large-scale empty space is scale-invariant.  
Under the above  key hypothesis, one is left with the following condition:
\begin{equation}
  \kappa_{\mu ;\nu}  + \kappa_{ \nu ;\mu}
 +2 \kappa_{\mu} \kappa_{\nu}
- 2 g_{\mu \nu} \kappa^{ \alpha}_{;\alpha}
 + g_{\mu \nu}\kappa^{ \alpha} \kappa_{ \alpha}   =  \lambda^2 \Lambda_{\mathrm{E}}  \, g_{\mu \nu}  .
\label{fcourt}
\end{equation}
\noindent
It is important to note that it has been verified in \citet{Maeder17a} that the line element in the scale-invariant empty space
is also conformally equivalent to the Minkowski metric.  Now, with the assumption that $\lambda$ is only a function of $t$, 
so that the cosmological principle holds, then one obtains the following non-zero terms $\kappa_0$ and $ \dot{\kappa_0}$ 
and by using Equation (\ref{k3}) one has $\kappa_0 \,= \, \dot{\lambda}/\lambda$, (dots indicating time derivatives).
Thus, the above condition (\ref{fcourt}) leads to
\begin{eqnarray}
\  3 \, \frac{ \dot{\lambda}^2}{\lambda^2} \, =\, \lambda^2 \,\Lambda_{\mathrm{E}}  \,  
 \quad \mathrm{and} \quad \frac{\ddot{\lambda}}{\lambda} \, = \,  2 \, \frac{ \dot{\lambda}^2}{\lambda^2} \, .
\label{diffd}
\end{eqnarray}

\noindent
These equations  establish a relation between the scale factor $\lambda$
and $\Lambda_{\mathrm{E}}$, which is the energy density of the empty space in GR.  At this point, it is important to 
recall that in GR, $\Lambda_{\mathrm{E}}$ and thus the 
properties of the empty space are  considered to not depend on the matter content of the Universe expressed
by $\Omega_{\mathrm{m}}$. We adopt the same assumption here.
This means that the above relations (\ref{diffd}) are always valid, whatever the matter content.  Evidently,  the same remark will apply for the solutions of these differential equations, {\it{i.e.}} for the expressions of the scale factor $\lambda(t)$.\\

\noindent
The solution of the above two equations (\ref{diffd})  is of the  form:
\begin{equation}
\lambda \, = \, \frac{y}{(t-b)^n} \, .
\label{l}
\end{equation}
\noindent
Both Equations (\ref{diffd}) imply $n=1$, while  $y \, = \, \sqrt{\frac{3}{\Lambda_{\mathrm{E}}}}$ follows from the first equation. 
Any value of the parameter $b$ would satisfy the equations. However, the value of $b$ has to be consistent 
with the cosmological equations  for the expansion factor $a(t)$ when applied to the limiting case of a zero density Universe, 
which have been derived previously, (29) - (31) in \citet{Maeder17a}, see Equations (\ref{EE1}), (\ref{EE2}) and (\ref{EE3}) below.
The results in \citet{Maeder17a}  clearly show that $b=0$ for zero density Universe. Indeed, there it has been considered the value of $b=0$ 
for the initial time; then it has been verified in Section 3.2 of \citet{Maeder17a} that the origin of the empty Universe, in this case $a(0)=0$, is at $t=0$. 

When the FLRW metric is used within the general field equations (\ref{field}), 
the three cosmological equations in the scale-invariant vacuum theory are:
\begin{equation}
\frac{8 \, \pi G \varrho }{3} = \frac{k}{a^2}+\frac{\dot{a}^2}{a^2}+ 2 \,\frac{\dot{a} \dot{\lambda}}{a \lambda} \, ,
\label{EE1}
\end{equation} 
\begin{equation}
-8 \, \pi G p  = \frac{k}{a^2}+ 2 \frac{\ddot{a}}{a}+\frac{\dot{a^2}}{a^2}
+ 4 \frac{\dot{a} \dot{\lambda}}{a \lambda}  \, .
\label{EE2}
\end{equation}
\begin{equation}
-\frac{4 \, \pi G}{3} \, (3p +\varrho)  =  \frac{\ddot{a}}{a} + \frac{\dot{a} \dot{\lambda}}{a \lambda}  \, .
\label{EE3}
\end{equation}
\noindent
In these equations, only two are independent. 
From the first equation, in the limit of zero density $\rho=0$ and flat space $k=0$ follows that
\begin{equation}
\frac{\dot{a}^2}{a^2}= -  2 \,\frac{\dot{a} \dot{\lambda}}{a \lambda}.
\end{equation}
\noindent
Using Equation (\ref{l}), the expression for the Hubble constant of the zero energy-density Universe is derived to be:
\begin{equation}
\frac{\dot{a}}{a}= \frac{2}{t-b} \, ,
\label{s0}
\end{equation}
\noindent
This gives an expansion factor $a(t)$:
\begin{equation}
a\, = \, \frac{a_{0}}{b^2} \, (t-b)^2 \, ,
\label{s1}
\end{equation}
where $a_{0}$ is a constant corresponding to the value of $a(t)$ at $t=0$. 
This solution also satisfies the other two equations  (\ref{EE3})  
in the limit of  flat space $k=0$ with zero pressure $p=0$ and density $\rho=0$.

The meaning of $b$ as ``the initial time of the Big Bang''  ($b=t_{in}$) is induced by considering  $a(t=b=t_{in})=0$.
If one chooses units such that the Big Bang is ``the beginning of the cosmic time'' as well, that is, at $t=b=t_{in}=0$, 
then Equation (\ref{s0}) gives $a(t)= a_0 t^2$,  for the empty space (see also Sect. 3.2 in \citet{Maeder17a}). 
Further, if the time units are set so that the current time epoch is chosen to be at $t=t_0=1$ with $a(t=t_0=1)=1$ 
then  $a_0=1$ and $t$ has the meaning of {\it normalized cosmic time} such that $a(t)=t^2$ with $t=t_0=1$ 
in the current time epoch and $t=0$ at the initial time of the Big Bang when $a(t=0)=0$. 
Finally, the normalized Hubble constant of the empty space is then $\frac{\dot{a}}{a}(t=t_0=1)=2$  and 
any deviations from this value are associated with the presence of matter. 
Thus, within the integrable Weyl geometry described by the Equations (\ref{field}), 
the corresponding FLRW Cosmological equations for empty space have a non-trivial solution for the scale factor $a(t)$. 
This solution in normalized cosmic time has a true Hubble constant of value 2.

Notice that in the case of the empty space there is an interesting interplay 
of the scaling factor $a(t)$ and the conformal factor $\lambda(t)$ which is best seen by considering normalized cosmic time $t$. 
While $a(t) \propto t^2$ grows with time from a zero value towards value of one in the present epoch, 
the conformal factor $\lambda(t) \propto t^{-1}$ shrinks from a huge infinite value to a finite value close to one.
In these coordinates the relevant FLRW metric now becomes $g'_{\mu\nu}=\lambda^2 g_{\mu\nu}$ 
with spatially constant scale factor since $\lambda^2 a(t)=y^2$ with $y \, = \, \sqrt{\frac{3}{\Lambda_{\mathrm{E}}}}$.
Thus, up to a conformal factor $y^2$, one has FLRW metric  $g'_{\mu\nu}$ with $a(t)=1$ at all times 
and the time-dependent component $g'_{00}=1/t^2$ can be turned into $g'_{00}=1$ upon the use of a logarithmic time 
$t\rightarrow\tau=\ln(t/t_0)$. In such  logarithmic time, the ``now'' is at $\tau=0$, the Big Bang is at $- \infty $, 
and the spacetime is locally Lorentz invariant since the metric tensor is now Minkowski like. 
The above considerations evidently apply to the empty space, thus the cosmological vacuum.

Therefore, one concludes that the scale factor of the empty space imposed by Equations (\ref{diffd}) is of the form $\lambda = y/t $.
This solution characterizes the properties of the macroscopic empty space assumed 
to be scale invariant  and with properties independent on the matter content of the Universe,  
the gauge $\lambda$ being fixed by the assumption that the macroscopic empty space is scale-invariant. 
To some extent, the expressions containing $\lambda$ replace the cosmological constant  $\Lambda_{\mathrm{E}}$
and preserve the scale invariance of the field equations. If $\lambda$ is chosen to be normalized as well; 
that is, $\lambda(t=t_0=1)=1$ then the constant $y\, = \, \sqrt{\frac{3}{\Lambda_{\mathrm{E}}}}$ 
has the meaning of ``age of the Universe'' or ``size of the Universe'' 
(note that units, where the speed of light is equal to 1, are in use in the above considerations) 
when the time scale of the current epoch is not such that $t_0=1$.\\

 \section{The equations of Euler, continuity, and Poisson in the scale-invariant context} \label{theory1}
 
Classically the equations of Euler and Poisson, as well as the continuity equation, 
form the basis for the study of the evolution of the density fluctuations in the Newtonian approximation. 
The first two equations are usually obtained in the weak field approximation of the general field equation in Riemann geometry,
now we consider the problem in the above mentioned context of the integrable Weyl's geometry.

 \subsection{The equation of motion} \label{euler}
 
The equivalent of the Newton equation in the scale-invariant framework 
was derived from the weak field approximation of the geodesic equation \citep{MBouvier79,Maeder17c}. 
This geodesic equation was itself derived from an action principle by 
\citet{Dirac73}, see also \citet{BouvierM78} for its interpretation in  the integrable Weyl's geometry.
The equation of motion in spherical coordinates is,
 \begin{equation} 
 \frac {d^2 \bold{r}}{dt^2} \, = \, - \frac{G \, M}{r^2} \, \frac{\bold{r}}{r} + \, \kappa(t)\, \frac{d\bold{r}}{dt} \, .
\label{Nvec}
\end{equation}
\noindent
The equation contains an additional term proportional to the velocity vector. Thus, in the case of an expansion of a system, 
there is an additional increase of the expansion;
in the case of an infall, the collapse is accelerated. The term $\kappa(t)$ is called the coefficient of metrical connection. 
As mentioned above,  it is usually a four-component field $\kappa_\nu$, however, 
due to considerations of isotropy and homogeneity of the 3D space only the time-dependent 
$\kappa_0$ component is non-zero.
Therefore, $\kappa_0$ is denoted by $\kappa(t)= - \dot{\lambda}/\lambda$ and thus $\kappa(t)= 1/t$. 
This means that the additional effects were larger at earlier epochs and that they are decreasing as time is passing.
In this context, as stated by \citet{MBouvier79}, the $g_{\mu \, \nu}$ and 
$\kappa_\nu$ terms, as well as the time $t$ are those of Special Relativity appropriate to the empty space. 
Here, the time $t$ is also  the time in a comoving system, that is to say, 
the cosmic time. Any observer attached to a comoving galaxy will measure the same age of the Universe.
This time starts at the Big Bang with $t_{in} =0$ and now it is 13.8 Gyr (different units may be considered). 

Interestingly enough, the above modified Newton equation, as well as the Euler equation, 
can also be derived in a rather simple way by the application of simple scale transformations.
From Equation (\ref{ds}), one also has 
\begin{equation}
r' \, = \lambda \; r \, , 
\end{equation}
\noindent 
since $\lambda$ is a function of time only, (as discussed above, symbols with a prime refer to the standard GR-framework, symbols without a prime 
are related to the scale-invariant framework). With the above relations, the velocities vectors in the two frameworks are related as follows:
 \begin{equation}
 \vec{v} \,'=\frac{d\vec{r}\, '}{dt'}=\frac{1}{\lambda}\frac{d(\lambda\vec{r})}{dt}=
 \vec{v}+\frac{\dot{\lambda}}{\lambda} \, \vec{r}=\vec{v}-\kappa \, \vec{r}\label{velocity}\, .
 \end{equation} 
\noindent
The operator $\vec{\nabla}$ transforms like $ \vec{\nabla}\,' = \vec{\nabla}/ \lambda$.

 \subsection{The equivalent of the Euler equation}

To derive the scale-invariant form of Euler equation consider the time derivatives (see the Appendix II in \citet{MBouvier79}),

\begin{eqnarray*}
\frac{d^{2}\vec{r}'}{dt'^{2}}=\frac{d}{dt'}\left(\frac{d\vec{r}'}{dt'}\right)=
\frac{1}{\lambda}\frac{d}{dt}\left(\frac{1}{\lambda}\frac{d}{dt}\left(\lambda\vec{r}\right)\right)=
\frac{1}{\lambda}\frac{d}{dt}\left(\dot{\vec{r}}-\kappa \vec{r}\right)=\\=
\frac{1}{\lambda}\left(\ddot{\vec{r}}-\kappa \dot{\vec{r}}-\dot{\kappa}\vec{r}\right) \,= 
\frac{1}{\lambda}\left(\ddot{\vec{r}}-\kappa \dot{\vec{r}}+\kappa^2\vec{r}\right) \, ,
\end{eqnarray*}
where $\kappa=-\dot{\lambda}/\lambda$ and $\ddot{\lambda}/\lambda=2\kappa^{2}$ have been employed,
(see Equation (25) in \citep{Maeder17a}) or equivalently that $\dot{\kappa}=-\kappa^{2}$.
The standard form of the Euler equation is:
\begin{equation}
\frac{d\vec{v}\,'}{dt'}=-\vec{\nabla}\,'\Phi'-\frac{1}{\rho'}\vec{\nabla}\,'p'+\frac{1}{3}\Lambda\vec{r\,'} \, .
\label{Eulp}
\end{equation}
\noindent
where for consistency with an equation of motion the $\Lambda_{\mathrm{E}}$ term \citep{Mavrides73} has been included as well. 
Here, the effect of the cosmological constant $\Lambda_{\mathrm{E}}$ within the general relativity framework has been included.
The new form of the Euler equation can be obtained by substituting
$\vec{v}'=\vec{v}- \kappa\vec{r}$ (\ref{velocity}) in the usual Euler
equation (\ref{Eulp}) along with the above mentioned scaling properties
for $r,\,\varrho,$ and $p$, using $\vec{\nabla}\lambda=0$,
and the scale invariance of the gravitational potential $\Phi'=\Phi$ (see Eq. (22) in \citet{Maeder17c}),
\[
\lambda\frac{d\vec{v}\,'}{dt'}=\left(\ddot{\vec{r}}-\kappa\dot{\vec{r}}+\kappa^2\vec{r}\right)=-
\vec{\nabla}\Phi-\frac{1}{\rho}\vec{\nabla}p+\frac{1}{3}\lambda^{2}\Lambda_{\mathrm{E}}\vec{r} \, ,
\]
\[
\ddot{\vec{r}}=-\vec{\nabla}\Phi-\frac{1}{\rho}\vec{\nabla}p+
\frac{1}{3}\Lambda_{\mathrm{E}}\lambda^{2}\vec{r}+\left(\kappa\dot{\vec{r}}-\kappa^2\vec{r}\right) \, .
\]
Using the expressions (22) and (23) in \citep{Maeder17a}, $\dot{\kappa}=-\kappa^{2}=-\lambda^{2}\Lambda_{\mathrm{E}}/3$,
one derives:
\[
\ddot{\vec{r}}=-\vec{\nabla}\Phi-\frac{1}{\rho}\vec{\nabla}p+\kappa^{2}\vec{r}+\left(\kappa\dot{\vec{r}}-\kappa^2\vec{r}\right)\, ,
\]
and by canceling the $\kappa^{2}$ terms the equation becomes:
\[
\frac{d\vec{v}}{dt}=-\vec{\nabla}\Phi-\frac{1}{\rho}\vec{\nabla}p+\kappa\dot{\vec{r}} \, .
\]
\noindent
Finally, the new Euler equation takes the form,
\begin{equation}
\frac{d\vec{v}}{dt}=\frac{\partial\vec{v}}{\partial t}+\left(\vec{v}\cdot\vec{\nabla}\right)\vec{v}=
-\vec{\nabla}\Phi-\frac{1}{\rho}\vec{\nabla}p+\kappa\vec{v}\label{Euler} \, .
\label{NvecE}
\end{equation}
\noindent
The equation of motion contains an additional term proportional to the velocity, as seen above in Equation (\ref{Nvec}). 
In the case of an outflow expansion, this term opposes the gravitational deceleration, 
while for an infall of matter, like for the growth of density fluctuations, the additional term contributes to an infall enhancement. 
This term plays a great role in the enhancement of density fluctuations in the scale-invariant theory.

\subsection{The equivalent of the continuity and Poisson equations } \label{obs}

Starting from the standard equations, where the symbols with primes are associated with the standard framework based on general relativity,
the equation of continuity and Poisson equations are:
\begin{equation}
\frac{\partial\varrho'}{\partial t'}+\vec{\nabla}'\cdot(\varrho'\vec{v'})=0\label{continuity'} \, ,
\label{contp}
\end{equation}
\begin{equation}
\vec{\nabla}'^{2}\Phi'={\triangle}'\Phi'=4\pi G\rho'\label{Poisson'} \, .
\label{Poissp}
\end{equation}
With the above transformations and scaling properties for $r'=\lambda r$, $\varrho=\lambda^2\varrho'$ 
and $p=\lambda^2 p'$ and $\vec{\nabla} \lambda=0$ the equation of continuity becomes:
\begin{equation} 
 \frac{1}{\lambda} \frac{\partial}{\partial t}\left(\frac{\varrho}{\lambda^{2}}\right)+
\frac{1}{\lambda}\vec{\nabla}\cdot\left(\frac{\varrho}{\lambda^{2}}\left(\vec{v}+\vec{r}\frac{\dot{\lambda}}{\lambda}\right)\right)=0\, ,
\end{equation}
\begin{equation}
\frac{1}{\lambda^3} \left[ \frac{\partial \rho}{\partial t} - 2 \frac{\dot{\lambda}}{\lambda}\rho+
\vec{\nabla}\cdot (\rho \vec{v})+ \frac{\dot{\lambda}}{\lambda} \vec{\nabla}\cdot (\rho \vec{r}) \right]=0 \,.
\end{equation}
\noindent
In the last term, the expression 
$ \vec{\nabla}\cdot(\rho \vec{r})= \vec{r}\cdot \vec{\nabla} \rho + \rho \vec{\nabla}\cdot \vec{r}= \vec{r} \cdot \vec{\nabla} \rho + 3 \, \rho $ 
can be used. Then the scale-invariant continuity equation becomes:
\begin{equation}
\frac{\partial \rho}{\partial t}+ \vec{\nabla}\cdot (\rho \vec{v}) = \kappa \left [\rho+ \vec{r} \cdot \vec{\nabla} \rho \right]\, .
\label{continv}
\end{equation}
\noindent
Evidently, if $\lambda$ is a constant ($\kappa=0$), the equation is the usual one. 
There is a term proportional to the density which in conventional treatments of such equations 
can be viewed as related to sinks and sources while
the second term is related to the density gradient and thus can be viewed as diffusion term from point of view of transport theory. 
However, it is important to remember that the additional terms come from the scaling, which was recalled above,
and are related to the corresponding conservation laws \citep{Maeder17a}.



Let us now turn to the Poisson equation. 
The potential $\Phi$ considered refers to the Newtonian contribution only and not to the additional term 
that the scale-invariant equation of motion is containing (see Sec. 2.2 by \citet{Maeder17c}). 
There, one has also seen the interesting property that the potential is an invariant, {\it{i.e.}} $\Phi'=\Phi$. 
Thus, (\ref{Poissp}) is now:
\begin{equation}
\frac{1}{\lambda^{2}}\triangle\Phi'=4\pi G\frac{\varrho}{\lambda^{2}} \, ,
\end{equation}
and noticeably the Poisson equation stays the same:
\begin{equation}
\vec{\nabla}{}^{2}\Phi=\triangle\Phi=4\pi G\varrho \, .
\label{Poisson}
\end{equation}
\noindent
For homogeneous density distribution $\rho$, the potential $\Phi$ and its gradient are ($\vec{\nabla} \varrho=0$):
$\Phi = \frac{2 \, \pi}{3} G \varrho r^2 , \; \mathrm{and} \; \vec{\nabla} \Phi = \frac{4 \pi}{3} G \varrho \vec{r} \, ,$
like in the usual Newtonian case.\\

\section{The application of the three equations to density perturbations in the scale-invariant theory} \label{theory2}

In the standard case, the theory has been developed by several authors, 
here the approach described by \citet{Theuns16} is followed.
In the expanding Universe, the position of a point is $\vec{r}=a\vec{x}$,
where $a=a(t)$ is the expansion factor and $\vec{x}$ is the co-moving
coordinate. Then the velocity of the considered point is 
\begin{equation}
\vec{v}=\dot{\vec{r}}=\dot{a}\vec{x}+a\dot{\vec{x}} \, ,
\end{equation}
\noindent
where the first term is the Hubble velocity, 
\begin{equation}
\dot{a}\vec{x}=Ha\vec{x}=H\vec{r}\, , \quad \mathrm{with} \quad H= \dot{a}/a \, .
\end{equation}
\noindent
 and $\vec{v}_{p}=a\dot{\vec{x}}$ is the peculiar velocity. 
Since functions can be expressed either via $\vec{x}$ coordinates
or via $\vec{r}$ then the corresponding differential operators will be:
\begin{equation}
\nabla=\frac{\partial}{\partial x}, \, \nabla a=0, \, \nabla_{r}=\frac{1}{a}\nabla,\, \nabla_{r}^{2}=\frac{1}{a^{2}}\nabla^{2},
\label{gradient}
\end{equation}
\begin{equation}
\left(\frac{\partial}{\partial t}\right)_{r}=\left(\frac{\partial}{\partial t}\right)_{x}-\left(H\vec{x}\cdot\vec{\nabla}\right)
\end{equation}
\noindent
This choice guarantees that $\left(\frac{\partial\vec{r}}{\partial t}\right)_{r}=0$
when $\vec{r}=a\vec{x}$ is used. Notice that all the previously discussed operators $\nabla$ were actually $\nabla_r$ type operators.

\subsection{The equation of continuity for the perturbation} \label{contper}

Now, consider a density function $\varrho$ related to the unperturbed ``background" density $\varrho_{b}$ and expressed as
\begin{equation}
 \varrho(\vec{x},t)\, = \, \varrho_{b}(t)(1+\delta(\vec{x},t)) \, ,
 \label{rho}
 \end{equation}
where $\delta(\vec{x},t)$ is the density perturbation. 
The continuity equation (\ref{continv}) is expressed in $\vec{\nabla}$ symbols with respect to $\vec{r}$.
By switching from $\vec{r}$ to $\vec{x}$ and inserting the above expression of $\varrho(\vec{x},t)$ in the 
continuity equation one obtains: 
\begin{eqnarray} 
\left(\left(\frac{\partial}{\partial t}\right)_{x}-H\vec{x}\cdot\vec{\nabla}\right)\varrho_{b}(t)(1+\delta)
+\frac{1}{a}\vec{\nabla}\cdot(\varrho_{b}(t)(1+\delta)(\dot{a}\vec{x}+\vec{v}_p) \\ \nonumber
=\kappa\left[\varrho_{b}(t)(1+\delta)+\vec{x}\cdot\vec{\nabla}\left(\varrho_{b}(t)(1+\delta)\right)\right] \, .
\end{eqnarray}
\noindent
By expanding the terms in this equation the expression becomes:
\begin{eqnarray}\label{long}
(1+\delta) \left(\frac{\partial\varrho_b}{\partial t}\right)_{x}
+\varrho_b \left(\frac{\partial\delta}{\partial t}\right)_{x} 
- H \vec{x}\cdot\vec{\nabla} (1+\delta)\varrho_{b}\\ \nonumber
+\frac{\dot{a}}{a}(1+\delta)\varrho_{b} \vec{\nabla} \cdot \vec{x}
+\frac{\dot{a}}{a}\vec{x}\cdot\vec{\nabla} (1+\delta)\varrho_{b} \\ \nonumber
+ \frac{\vec{v}_p}{a} (1+\delta) \cdot \vec{\nabla} \varrho_b 
+ \frac{\varrho_b}{a} \vec {\nabla} \cdot (1+\delta) \vec{v}_p\\ \nonumber
=\kappa\left[ (1+\delta)\varrho_b +\vec{x} \cdot \vec{\nabla} \varrho_b(1+\delta) \right] \, .
\end{eqnarray}
\noindent
The third and fifth terms cancel while the sixth term is zero since the background density is homogeneous.
\begin{eqnarray}\label{short}
(1+\delta) \left(\frac{\partial\varrho_b}{\partial t}\right)_{x}
+\varrho_b \left(\frac{\partial\delta}{\partial t}\right)_{x} 
+\frac{\dot{a}}{a}(1+\delta)\varrho_{b} \vec{\nabla} \cdot \vec{x}
+ \frac{\varrho_b}{a} \vec {\nabla} \cdot (1+\delta) \vec{v}_p\\ \nonumber
=\kappa\left[ (1+\delta) \varrho_b+\vec{x} \cdot \vec{\nabla}\delta\varrho_b\right] \, .
\end{eqnarray}
If one considers homogeneous background density and only the terms independent of the perturbation, 
that is to say, independent of $\delta$ and $\vec{v}_p$, then one gets:
\begin{equation}
\dot{\varrho}_b+H \varrho_b \vec{\nabla}\cdot \vec{x} = \kappa \varrho_b \, \quad \mathrm{or} \; \; 
\dot{\varrho}_b+3 \,H \varrho_b = \kappa \varrho_b \, .
\label{rob}
\end{equation}
\noindent
This equation expresses the evolution of the background density. In the standard case, the right-hand side is zero.
The same relationship, but multiplied by $\delta$, is contained in Equation (\ref{short}). 
This allows further simplifications in this equation, which now only remains as:
\begin{equation}
\varrho_b \left(\frac{\partial\delta}{\partial t}\right)_{x} 
+ \frac{\varrho_b}{a} \vec{\nabla} \cdot (1+\delta) \vec{v_p} = 
\kappa \vec{x} \cdot \vec{\nabla}\varrho_b \delta \, ,
\end{equation}
\noindent
where the derivative of $\delta$ is at constant $x$.
With further manipulations using that $\varrho_b$ is homogeneous and the expression 
$\vec{v_p} = a \, \vec{\dot{x}}$, this equation finally becomes:
\begin{equation} 
\dot{\delta}+ (1+\delta)\vec{\nabla} \cdot \dot{\vec{x}} = 
\kappa \, \vec{x} \cdot \vec{\nabla}\delta \, ,
\label{contdelta}
\end{equation}
where the partial time derivative $\partial\delta/{\partial t}$ has been combined with the term 
$\dot{\vec{x}}  \cdot \vec{\nabla}\delta $ to the full time derivative $\dot\delta$.

Let us examine this equation, which depends on the density gradient  $\vec{\nabla} \delta$ of the perturbation.
The product $\vec{x} \cdot \vec{\nabla}\delta$ expresses the projection of the density gradient along the vector $\vec{x}$, 
corresponding to  the  displacement of a piece of the fluid element initially  at rest in the comoving medium.
Indeed, one is considering a particular element of the background medium of density $\varrho_b$: from its initial position,
this fluid element starts moving with a velocity $\vec{v_p}$ towards the nascent density fluctuation considered as spherically symmetric.
Thus, the product  $\vec{x} \cdot \vec{\nabla}\delta$
representing the above projection is positive, since the directions of both vectors  are the same.  Therefore, one has:
\begin{equation}
\vec{x} \cdot \vec{\nabla}\delta \approx x \frac{\partial \delta}{\partial x}
= \frac{\partial \ln \delta}{\partial \ln x} \, \delta\, \equiv \, n \, \delta \,.
\label{dlog}
\end{equation}
\noindent

Alternatively, one can consider separation of the spatial and time dependence of $\delta$ 
such that $\delta(t,\vec{x})=\delta(t)\delta(\vec{x})$.  
The assumption of separability applies only to pressureless perturbations, which is the case considered in this paper.
Then $n$ can be defined as the order of homogeneity according to the Euler's theorem for homogeneous functions 
($\vec{x}\cdot\nabla\delta(\vec{x})=n\delta(\vec{x})$). 
Thus, $n$ encodes information only about the geometric shape of $\delta(\vec{x})$ while the magnitude of $\delta(\vec{x})$ is arbitrary.
From what follows, the anticipated typical value is $n=2$, but in Sec. \ref{num} 
the effects of nearby values such as $n= 1$, 3 and 5 will be explored as well. 
The linearized equation for the density fluctuation finally becomes:
\begin{equation}
\dot{\delta}+ (1+\delta) \vec{\nabla} \cdot  \dot{\vec{x}} = n \,\kappa(t) \, \delta \, .
\label{contdelta}
\end{equation}
This equation shows that positive values of $n$, as indicated above, lead to a growth of the density fluctuations.
In the standard case, the right-hand side is zero. Thus, scale invariance leads to an additional growth 
of the density fluctuations, which depends on the slope $n$ of the density gradient. 
The amplitude of this term is also dependent on the epoch because of $\kappa=1/t$.

The outer layers of clusters of galaxies may be represented by an isothermal polytrope, {\it{i.e.}} with a polytropic index equal to 5. 
The distribution of the density in the outer layers of such a polytrope follows a law of the form $\varrho \sim 1/r^2$ \citep{Chandra60}. 
This density profile has been confirmed by extended numerical simulations, leading to the so-called NFW profile, which in the outer layers
of a cluster is very close to the $\varrho \sim 1/r^2$ profile \citep{NFW95}. Within the core of a cluster at equilibrium, 
the density law is slightly less steep with a density law between laws with 
$r^{-2}$ and $r^{-1}$ \citep{NFW95,Ettori00}. This central profile corresponds to the equilibrium
structure, which is not necessarily fully reached during the relevant formation process under study. 
These density profiles are homogeneous functions of order $-2$ and $-1$. 
An extreme example of  homogeneous function of order $-3$ is a density profile described by Dirac delta function.
Such distributions with a negative order of homogeneity index appear to be in contradiction with the 
conclusion based on (\ref{dlog}) that $n$ must be positive in order to observe some growth of the density fluctuations. 
This raises a question about the actual value of $n$ that commands the need of understanding of  (\ref{dlog}) better.

First, it is very likely that there is a spectrum of possible density fluctuation types with various values of $n$ 
that can evolve as the Universe unfolds. 
Preliminary numerical calculations show that even if one starts and runs the numerical integration with a fixed $n$, 
fluctuations with other values of the order of homogeneity index $n$ enter $\delta$ since $\dot{\delta}/(\kappa(t) \, \delta)$ 
is not an integral of the motion. However, $\dot{\delta}/(\kappa(t) \, \delta)$ evolves 
and always settles at a positive value that stays constant for a significant fraction of the evolution.
This is true even if one considers negative values for $n$ ($n<0$) within the relevant evolution equations.

Second, to understand how starting with $\dot{\delta}/(\kappa(t) \, \delta)<0$ the system can evolve into
$\dot{\delta}/(\kappa(t) \, \delta)>0$ one has to look at the term $\vec{\nabla} \cdot  \dot{\vec{x}}$.
The gradient $\vec{\nabla}$ is pointing outwards from the center of the coordinate system 
($\vec{\nabla}\cdot\vec{x}=3$) where the density fluctuation is centered.
The background matter, with density $\varrho_b$, normally follows Hubble expansion. 
Then consider a particular element of this background fluid. 
From its initial comoving position, this fluid element starts moving with a particular velocity 
$\vec{v}_{\mathrm{p}}$ towards the nascent density fluctuation. 
Thus the direction of $\vec{v}_{\mathrm{p}}$ is pointing towards the origin of the coordinate system 
and therefore $\vec{\nabla} \cdot  \dot{\vec{x}}<0$ since the direction of the two vectors are opposite.
When this term $\vec{\nabla} \cdot  \dot{\vec{x}}$ is moved from the left-hand side 
to the right--hand side of (\ref{contdelta}) it contributes a positive factor to $\dot{\delta}$. 
This positive factor can overwhelm a negative term $n\kappa(t)\delta$ when $n<0$ and 
produce an effective $n$ that is positive $\dot{\delta}/(\kappa(t) \, \delta)>0$ as needed
for the growth of the density fluctuations.

In the above context, a spherical perturbation is considered with a radial infall motion directed towards the center. 
Thus, considering $0< n=1,2,3 $ is well justified, especially if one considers 
spherical bubble nucleation as the likely density fluctuation in the early universe. 
Such bubble nucleation can be induced by the matter-antimatter annihilation processes and 
the phase transitions in the very early universe that result in lower local matter density 
than the surrounding of the annihilation event \citep{Marques92,Liu92}.
Furthermore, since the goal is to study the growth of structure in the Universe, then this is 
about local near equilibrium confinement of a system. Such near equilibrium states
are usually well described by local harmonic oscillator potential energy ($\approx \frac{\omega^2}{2} r^2$).  
Thus, the above considerations justify the special place of $n=2$ in the forthcoming discussions.

\subsection{The Poisson equation for the perturbation} \label{Poissper}
 
It has been discussed already that in the scale-invariant context the Poisson equation remains the same, see Equation (\ref{Poisson}).
The density perturbations influence the Newtonian potential $\Phi$. 
With relation (\ref{gradient}), the Poisson equation for the perturbed density becomes:
\begin{equation}
\frac{1}{a^{2}}\vec{\nabla}{}^{2}\Phi=4\pi G\varrho_{b}(1+\delta)
\end{equation}
\noindent
Let's define a potential $\Psi$ due to the perturbation of density:
\begin{equation}
\vec{\nabla}{}^{2}\Psi=4\pi Ga^{2}\varrho_{b}\delta \, .
\label{Psidelta}
\end{equation}
\noindent
Thus, the total potential $\Phi$ may be expressed as:
\begin{equation}
\vec{\nabla}{}^{2}\Phi= \vec{\nabla}{}^{2}\Psi+4\pi G\varrho_{b}a^2 \, .
\label{Poper}
\end{equation}
\noindent
Up to a sign, the Newtonian force due to the total potential becomes:
\begin{equation}
\vec{\nabla}\Phi=\vec{\nabla}\Psi+\frac{4\pi}{3}G\varrho_{b}a^{2}\vec{x}
\end{equation}
\noindent
It will be useful to have a specific potential $\Psi$ attached to the perturbation of the background density in view of the 
Euler equation describing the motions associated with perturbations.

\subsection{The Euler equation for the perturbation in the expanding Universe}

In an expanding Universe with perturbed density fluctuation the scale-invariant equation 
corresponding to the Euler equation (\ref{NvecE}) was derived in the $r$ coordinates:
\begin{eqnarray}
\frac{d\vec{v}}{dt}=\frac{\partial\vec{v}}{\partial t}+\left(\vec{v}\cdot\vec{\nabla_r}\right)\vec{v}=
-\vec{\nabla_r}\Phi-\frac{1}{\varrho}\vec{\nabla_r}p+\kappa\vec{v}\label{Euler} \, .
\end{eqnarray}
After taking care that the derivative $\frac{\partial}{\partial t}$ is taken at constant $\vec{x}$,
and by re-expressing it in the comoving $\vec{x}$ coordinates and including the terms of the potential $\Psi$,
the Euler equation becomes:
\begin{eqnarray}
\left(\frac{\partial}{\partial t} -H \vec{x}\cdot\vec{\nabla}\right)(\dot{a} \vec{x}+\vec{v}_p)+
\frac{1}{a} [(\dot{a} \vec{x}+\vec{v}_p)\cdot \vec{\nabla}](\dot{a} \vec{x}+\vec{v}_p)= \\ \nonumber
- \frac{1}{a} \left(\vec{\nabla} \Psi+ \frac{4 \pi}{3} G \varrho_b a^2 \vec{x} \right)+\kappa(t)(\dot{a} \vec{x}+\vec{v}_p)-
\frac{1}{a \varrho} \vec{\nabla} p \, .
\end{eqnarray}
\noindent 
By expanding the various terms in the above equation and considering a polytropic gas $p=p(\varrho)$, 
where $\vec{\nabla} p = v^2_s \, \vec{\nabla} \varrho$, and $v_s$ is the sound velocity, one gets:
\begin{eqnarray}
\label{long2}
\ddot{a} \vec{x} +\dot{\vec{v}}_p - H \dot{a}\vec{x}- H (\vec{x}\cdot\vec{\nabla})\vec{v}_p+
\frac{1}{a} \left[ \dot{a}^2 \vec{x}+\dot{a}(\vec{x} \cdot \vec{\nabla}) \vec{v}_p+\dot{a} \vec{v}_p +
(\vec{v}_p \cdot \vec{\nabla}) \vec{v}_p \right]= \\ \nonumber
- \frac{1}{a} (\vec{\nabla} \Psi+ \frac{4 \pi}{3} G \varrho_b a^2 \vec{x}) - \frac{v^2_s}{a} \, \frac{\vec{\nabla} \varrho}{\varrho}+
\kappa(t) (\dot{a} \vec{x}+\vec{v}_p) \, .
\end{eqnarray}
\noindent
On the left side, the third term cancels with the fifth one, the fourth term cancels with the sixth one. 
Note that the scale-invariant cosmological models obey the following equation in the limit $p=0$
(see Eq. (\ref{EE3})  which is also Equation (31) in \citet{Maeder17a}):
\[
- \frac{4 \pi G \varrho_b}{3} = \frac{\ddot{a}}{a}+ \frac{\dot{a}}{a} \frac{\dot{\lambda}}{\lambda} \, , \quad \mathrm{or}
\quad \ddot{a}= - \frac{4 \pi G \varrho_b a}{3}+ \dot{a} \kappa(t) \, .
\]
 \noindent
Such equation can also be obtained by setting $v_p=0$ and demanding that the resulting equation is true for any $\vec{x}$. 
This brings further simplifications in Equation (\ref{long2}) where the terms 
$\kappa(t) \dot{a} \vec{x}$ and $\ddot{a} \vec{x}$ on both sides will cancel as well,
 \begin{equation}
 \dot{\vec{v}}_p+ H \vec{v}_p+ \frac{1}{a}(\vec{v}_p \cdot \nabla) \vec{v}_p=- \frac{\vec{\nabla} \Psi}{a}-
 \frac{v^2_s}{a} \, \frac{\vec{\nabla} \varrho}{\varrho} +\kappa(t) \vec{v}_p \, .
 \end{equation}
\noindent
By using $\vec{v}_p = a \dot{\vec{x}}$ and $ \dot{\vec{v}}_p = \dot{a} \dot{\vec{x}}+ a \ddot {\vec{x}}$ this equation becomes:
\begin{equation}
\ddot{\vec{x}}+ 2 H \dot{\vec{x}}+ (\dot{\vec{x}}\cdot \vec{\nabla}) \dot{\vec{x}} \, = \,
- \frac{\vec{\nabla} \Psi}{a^2}- \frac{v^2_s}{a^2} \, \frac{\vec{\nabla} \varrho}{\varrho} +\kappa(t) \dot{\vec{x}} \, .
\label{Eulerp}
\end{equation}
\noindent
This is the expression of the Euler equation applied to the perturbation of the density in the comoving $\vec{x}$ coordinates system. 
It differs from the standard case by the last term on the right. In the above expression, the pressure term is indicated. 
However, we have already discussed that the approximation made in the equation of continuity is valid for pressureless perturbations. 
Therefore in what follows we will consistently consider the perturbations in a dust universe where the  pressure term is absent.

\subsection{The evolution of the density contrast in the scale-invariant theory} \label{evoldelta}

The three equations (\ref{contdelta}), (\ref{Poper}), and (\ref{Eulerp}) form the necessary basis for 
the study of the evolution of perturbation in a dust Universe. In more general cases,
for example for the radiative era, an energy equation must be added \citep{Peebles80,Theuns16}. The scale-invariant effects appear through
the terms containing $\kappa(t)$, which are present in Equations (\ref{contdelta}) and (\ref{Eulerp}) but not in (\ref{Psidelta}).
Let's concentrate on small perturbations and examine the linear approximation of these three equations.
Under this hypothesis, Equation (\ref{contdelta}) may be simplified:
\begin{equation}
\vec{\nabla}\cdot\vec{\dot{x}}
=\frac{1}{(1+\delta)}\left(n \, \kappa(t) \delta-\dot{\delta}\right)
\approx n \, \kappa(t) \delta-\dot{\delta}+O(\delta^2)
\label{graddotx}
\end{equation}
\noindent
Thus, $\vec{\nabla}\cdot\vec{\dot{x}}$ is at least first order in $\delta$.
Neglecting terms of order higher than on $\delta$, the continuity equation for the perturbation simplifies further. 
Finally, in a dust universe, the three equations become:
\begin{equation}
\dot{\delta} + \vec{\nabla} \cdot \dot{\vec{x}} \, = \, n \, \kappa(t) \delta \, ,
\label{dn}
\end{equation}
\begin{equation}
\vec{\nabla}{}^{2}\Psi=4\pi Ga^{2}\varrho_{b}\delta \, ,
\end{equation}
\begin{equation}
\ddot{\vec{x}}+ 2 H \dot{\vec{x}}+ (\dot{\vec{x}}\cdot \vec{\nabla}) \dot{\vec{x}}\, = \,
- \frac{\vec{\nabla} \Psi}{a^2} +\kappa(t) \dot{\vec{x}} \, .
\label{Eulerd}
\end{equation}
Now, consider the time derivative of (\ref{dn}) and take the divergence of (\ref{Eulerd}) to obtain:
\begin{equation}
\ddot{\delta} + \vec{\nabla} \cdot \ddot{\vec{x}} \, = \, n \, \dot{\kappa} \delta + n \kappa \dot{\delta} \, ,
\label{ddelta3}
\end{equation}

\begin{equation}
\vec{\nabla} \cdot \ddot{\vec{x}}+ 2 H \vec{\nabla} \cdot \dot{\vec{x}}+\vec{\nabla} \cdot(\dot{\vec{x}}\cdot \vec{\nabla}) \dot{\vec{x}}\ \, = \,
- \frac{\vec{\nabla^2} \Psi}{a^2} +\kappa(t) \vec{\nabla} \cdot \dot{\vec{x}} \, .
\label{E3}
\end{equation}
By using (\ref{graddotx}) and similar reasoning as in (\ref{graddotx}) to justify neglecting 
$\vec{\nabla} \cdot(\dot{\vec{x}}\cdot \vec{\nabla}) \dot{\vec{x}}\approx O(\delta^2)$, one has:
\[
\vec{\nabla} \cdot \ddot{\vec{x}}+ 2 H \, (n \, \kappa\, \delta-\dot{\delta})\, = \,
- \frac{\vec{\nabla^2} \Psi}{a^2} +\kappa\, (n \, \kappa\, \delta-\dot{\delta}) \, .
\]
By subtracting this equation from (\ref{ddelta3}), one arrives at: 
\begin{equation}
\ddot{\delta}+ (2H -(1+n) \kappa)\dot{\delta} \, =
 \, 4\pi G \varrho_{b}\delta + 2 n \kappa (H-\kappa) \delta \, .
 \label{D2}
 \end{equation}
 \noindent
 This is the basic equation for the growth of density perturbations in the scale-invariant theory. 
 The term $\kappa$ multiplying $\dot{\delta}$ originates both from the continuity and Euler equations. On the right, the term
 $\kappa$ multiplying $H$ comes from the continuity equation, while the products $\kappa^2$ comes from both equations.
 The terms with $\kappa= - \dot{\lambda}/\lambda$ are absent in the standard models (there $\lambda=1$) where one 
 has the well-known equation:
 \begin{equation}
\ddot{\delta}+ 2H \dot{\delta} \, = \, 4\pi G \varrho_{b}\delta \, .
 \label{D3}
 \end{equation}
 \noindent 
 For the standard EdS model of a matter dominated Universe, where $a(t) \propto t^{2/3}$, this last equation predicts 
 a growth of the perturbation $\delta \propto t^{2/3} \propto a(t) \propto (1+z)^{-1}$. Thus, starting from the perturbation of the CMB
 $\delta \approx 10^{-5}$ at a redshift $z \approx 10^{3}$, the density fluctuations should be of the order of $10^{-2}$ at the present
 time. This is clearly much smaller and thus in strong contradiction with the observed density structures in today's Universe, a fact which is often
 considered as one of the most convincing argument in favor of the existence of dark matter \citep{Theuns16}. 
 
\subsection{The coefficients in the equation for the density perturbations} 
 
\begin{figure*}[tbh]
\centering
\includegraphics[width=.55\textwidth]{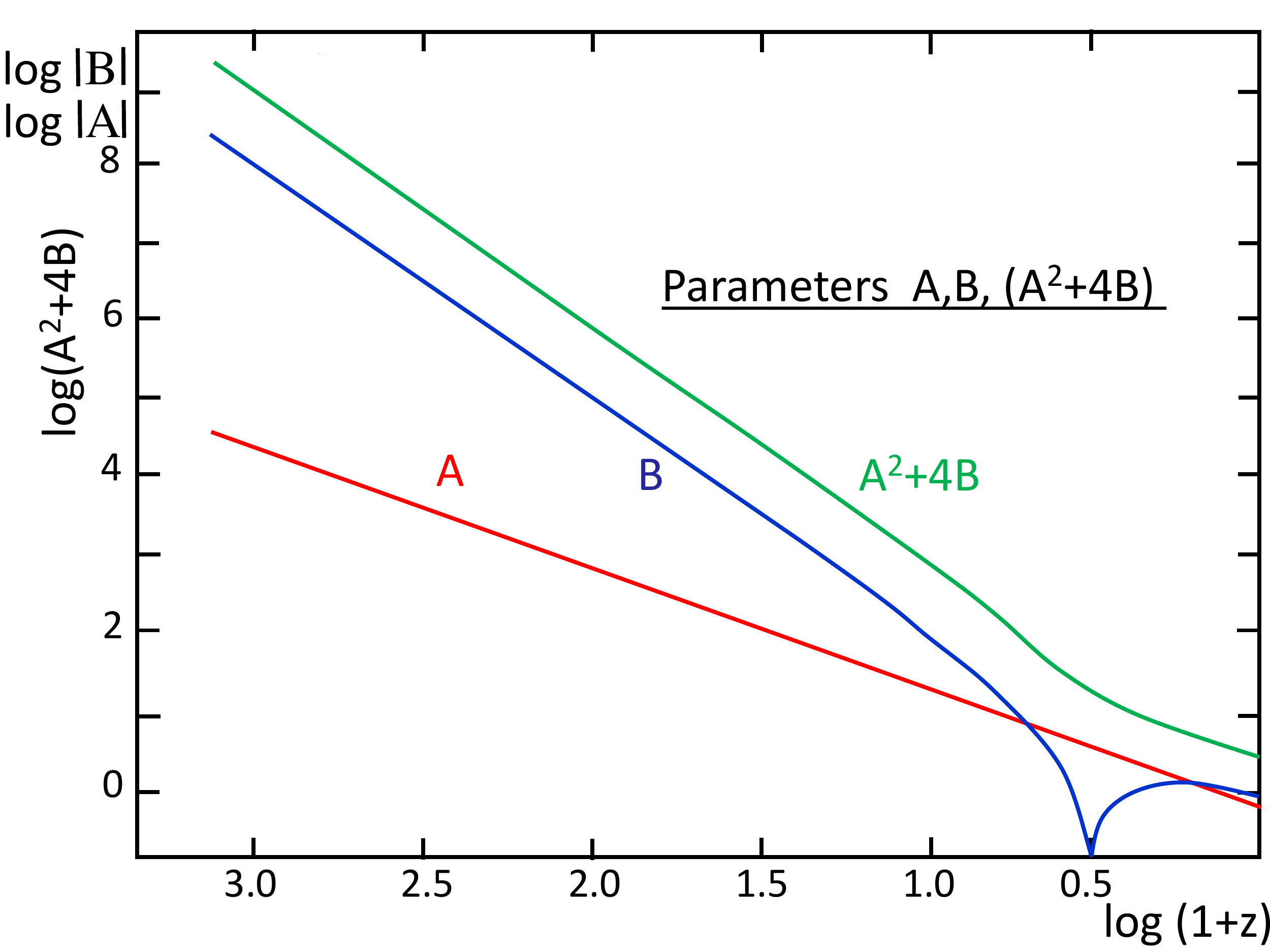}
\caption{The redshift dependence of the parameters $A$ and $B$ for the model with $\Omega_{\mathrm{m}}= 0.10$ and $n=2$. 
Note that $A$ is always negative, $B$ is negative for $\log (z+1) > 0.50$ and positive for smaller redshifts. 
The sum ${A^2 + 4 B}$ is always positive so that the discriminant $\sqrt{A^2 + 4 B}$ 
of the differential equation (\ref{D4}) is positive for all relevant redshifts.}
\label{AB}
\end{figure*}

The differential equation (\ref{D2}), which determines the evolution of density perturbations in the scale-invariant context, 
is a homogeneous differential equation of the second order with non-constant coefficients. It may be written as
 \begin{equation}
\ddot{\delta}+ A \, \dot{\delta} \, = \, B \, \delta \, ,
 \label{D4}
 \end{equation}
 \begin{equation} 
 \mathrm{with} \quad A= 2H -(1+n) \kappa \quad \mathrm{and} \quad B=4 \pi G \varrho_{b} + 2 n \kappa (H-\kappa) \, . 
 \end{equation}
 \noindent 
The coefficient $A$ expresses the damping factor. In the standard case, where $A= 2 H$, 
 this term contributes negatively to the second derivative $\ddot \delta$, thus acting as a damping factor. 
 The reason is that the Hubble expansion gives an opposing effect to the accumulation of matter on a particular point of space;
 thus, it does not favor the growth of density fluctuations. As a consequence, the growth of the density 
 perturbation in the standard case of an expanding universe tends to follow a power law instead 
 of an exponential growth as in the static case. In the scale-invariant case, the additional acceleration term 
 in Equations (\ref{Nvec}) and (\ref{NvecE}) also favors a power law, however with a much larger power than in the standard EdS case.
 If coefficient $A$ is negative, which is generally the case as seen in Fig. \ref{AB}, it acts as a growth factor.
 There the additional $\kappa$ term contributes to increase the infall and this effect overcomes the opposing effect
 of the Hubble expansion during most of the evolution (Fig. \ref{Hk}).

The coefficient $B$ contributes, if positive, to the growth of the density fluctuations. The term $4 \pi G \varrho_{\mathrm{m}}(t)$
expresses the effect of the Newtonian acceleration which always favors the accumulation of matter for a positive density
fluctuation. The second term results from the effects of the continuity and Euler equations. Its contribution depends
on the sign of $H-\kappa$, which in a given model changes according to the epochs as shown in Fig. \ref{Hk}. 
From this figure, one can see that during most of the  evolution  of the density perturbations, the contribution 
of the $H-\kappa$ term is negative.

Since the coefficients are not constant, there is no general expression for the solutions of Equation (\ref{D4}). 
As stated in Sec. \ref{euler}, the time $t$ in the modified Euler equation and in the continuity equation is the cosmic time, 
i.e the time in co-moving galaxies. Also, note that the above choice is also consistent with a timescale of 
the empty scale-invariant cosmological model that starts at time $t=0$. 
One should also notice that these equations are ``gauge invariant"  as long as one 
takes care to use the correct functional forms of the corresponding quantities, 
like $H$, $\kappa$ and so on for the chosen time gauge.
 
For the purpose of the easier comparison with observations, however, the constants $A$ and $B$ and their ingredients 
$H$, $\kappa$ and $4 \pi G \varrho_{b}$ have to be consistently expressed in the cosmic time $t$, which is different from the 
arbitrary timescale, say $\tau$, used in the early applications of the scale-invariant cosmological models.
The expression giving the cosmic time $t$ as a function of $\tau$ is: 
\begin{equation}
 t \, = \, \frac{\tau-\tau_{\mathrm{in}}}{\tau_0 - \tau_{\mathrm{in}}}\Delta_0 \, , \quad \mathrm{with}
\quad \tau_{\mathrm{in}}= \tau_0 \, \Omega^{\frac{1}{3}}_{\mathrm{m}} \, ,
\label{tt}
\end{equation}
where $\Omega_{\mathrm{m}}= \varrho/\varrho_{\mathrm{c}}$ is the usual density parameter with $\varrho_{\mathrm{c}}=3 H^2_0/(8 \pi G)$. 
The present time $\tau_0$ is fixed to $\tau_0=1$ in dimensionless time units $\Delta_0=1$ ($\tau$-scale). 
To get the cosmic time in years, $\Delta_0$ may be taken equal to $13.8 \cdot 10^9$ yrs. \citep{Frie08}. 
The analytical expressions of the scale factor $a(\tau)$ and $H(\tau)$ for flat scale-invariant models have been given by \citet{Jesus17}, 
they exactly correspond to the numerical solutions developed in cosmological models by \citet{Maeder17a}. In the $\tau$-scale, 
the age of the universe is $\tau_{\mathrm{tot}}= 1-\tau_{\mathrm{in}} = 1- \Omega^{\frac{1}{3}}_{\mathrm{m}} $. 
To express the Hubble ``constant" in the cosmic timescale $t$ consider the following relations:

\begin{figure*}[t!]
\centering
\includegraphics[width=.55\textwidth]{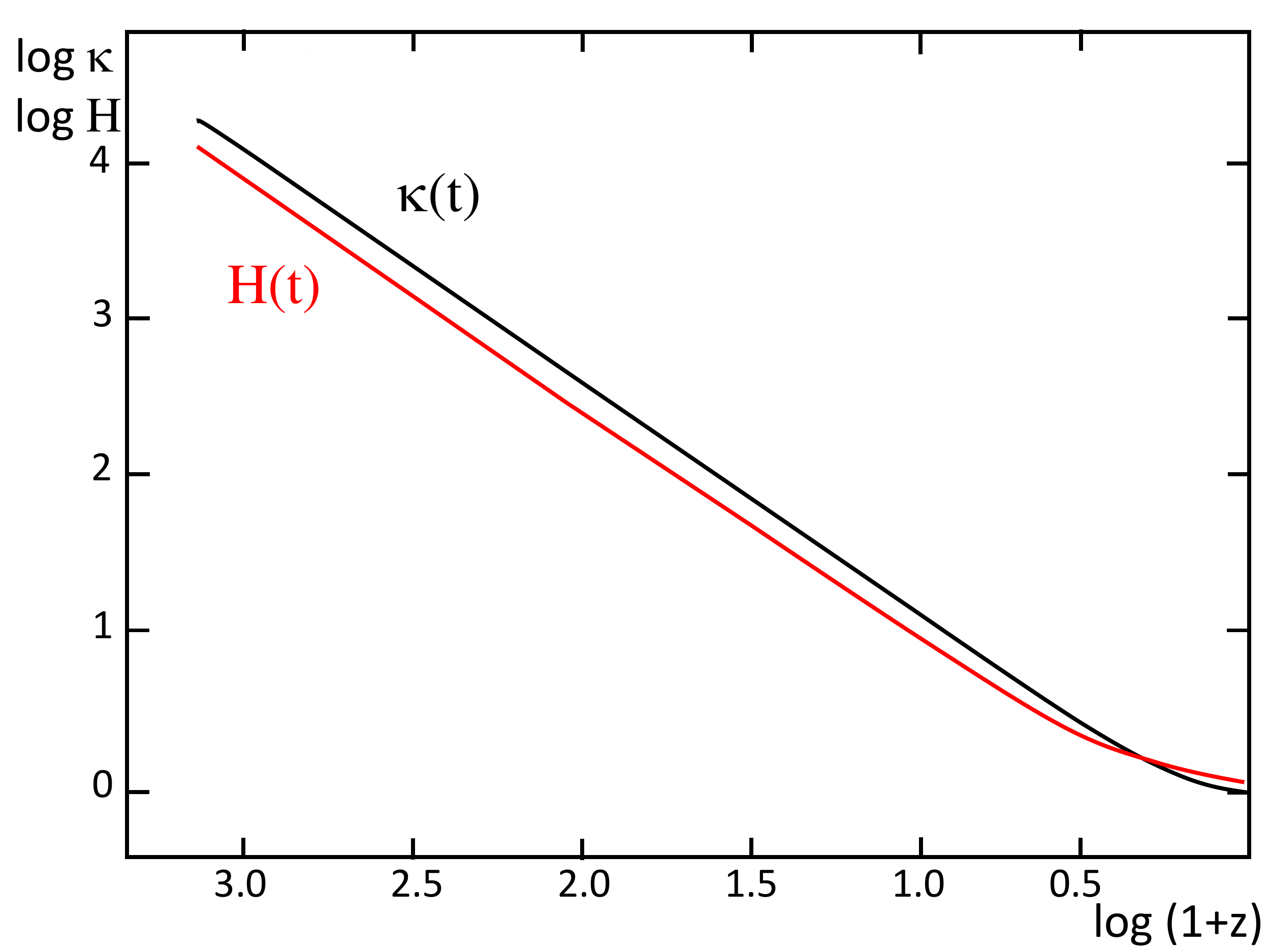}
\caption{Example of the numerical values of the Hubble constant $H$ and the parameter $\kappa=1/t$ in the cosmic timescale $t$
as a function of the redshift for the case $\Omega_{\mathrm{m}}=0.10$. To get the Hubble constant 
in [km/(s Mpc)] at various redshifts, the values of $H$ have to be multiplied by relevant factor $\Delta_0^{-1}$, about 70 [km/(s Mpc)].}
\label{Hk}
\end{figure*}

\begin{equation}
H(t) \, = \,\frac{1}{a}\frac{da}{\,dt} = \frac{1}{a}\frac{da}{\,d\tau} \frac{d \tau}{dt} = H(\tau) \frac{d \tau}{dt}=
 H(\tau) (1- \Omega^{\frac{1}{3}}_{\mathrm{m}})\times\Delta_0^{-1} \, , \quad \mathrm{with} \; \;
 H(\tau) = \frac{2 \tau^2}{\tau^3- \Omega_{\mathrm{m}}}\, ,
 \label{hh}
\end{equation}
\noindent
according to the expression of $H(\tau)$ given by \citet{Jesus17}. 
To evaluate the Hubble constant in [s$^{-1}$] units at a given epoch, 
the value of $H(t)$ in Equation (\ref{hh}) should be divided by the current age of the universe $\Delta_0=4.355\times10^{17}$ s.
There is also a hidden timescale in the term $4 \pi G \varrho$, since $\sqrt{G \, \varrho}$ is the inverse of the oscillation timescale of the system of density $\varrho$. For a density parameter $\Omega_{\mathrm{m}}$ the present-day density of the Universe writes as:
\begin{equation}
\varrho_{\mathrm{m}}(t_0) = \Omega_{\mathrm{m}} \, \varrho_{\mathrm{c}}(t_0) =
 \Omega_{\mathrm{m}} \, \frac{3 \, H^2_0(t_0)}{8 \pi G}= 
 \Omega_{\mathrm{m}} \, \frac{3 \, H^2_0(\tau_0)}{8 \pi G} (1- \Omega^{\frac{1}{3}}_{\mathrm{m}} )^2 \times \Delta_0^{-2}\, . 
 \label{rr}
 \end{equation}
 \noindent
These expressions in (\ref{rr}) and (\ref{hh}) show that the Hubble ``constant" $H$ behaves as co-scalar of rank $-1$, 
while the density $\varrho$ behaves as co-scalar of rank $-2$; furthermore, based on (\ref{D2}) one see that $\delta$ is a in-scalar
according to the terminology recalled  in Section \ref{base}.

Since in the scale-invariant cosmological models the matter density obeys the conservation law 
$\varrho_{\mathrm{m}}(\tau) a^3(\tau) \lambda(\tau)= const.$ as shown by Equation (61) in 
\citep{Maeder17a}, then 
$\varrho_{\mathrm{m}}(\tau) a^3(\tau) \lambda(\tau)=\varrho_{\mathrm{m}}(\tau_0) a^3(\tau_0) \lambda(\tau_0)=
\varrho_{\mathrm{m}}(\tau_0) \lambda(\tau_0)$, 
because $\lambda(\tau_0)/\lambda(\tau)=\tau$ and $a(\tau_0)=1$ in the $\tau$-scale one gets:
\begin{equation} 
4 \pi G \varrho_{\mathrm{m}}(\tau) = \frac{4 \pi G \varrho_{\mathrm{m}}(\tau_0) \, \tau}{ a(\tau)^3} \,
\label{ppi}
\end{equation}
\noindent
This result is consistent with the integration of (\ref{rob}) using the exact solution by \citet{Jesus17}
when the evolution of the background density in the cosmological models has been calculated within the $\tau$-scale.
Then by using $ \varrho_{\mathrm{m}}(t)=\varrho_{\mathrm{m}}(\tau)(1- \Omega^{\frac{1}{3}}_{\mathrm{m}} )^2 \times \Delta_0^{-2}\,$ 
the term $4 \pi \varrho_{\mathrm{m}}(t)$ at a certain cosmic time $t$ can be written as:
\begin{equation} 
 4 \pi G \varrho_{\mathrm{m}}(t) = 
 4 \pi G \, \Omega_{\mathrm{m}}\varrho_{\mathrm{c}}(\tau_0) (1- \Omega^{\frac{1}{3}}_{\mathrm{m}} )^2 
 \times \Delta_0^{-2}\,\frac{\tau(t)}{a^3(\tau(t))}=
 \frac{3}{2} \Omega_{\mathrm{m}}H^2_0(\tau_0) \frac{ (1- \Omega^{\frac{1}{3}}_{\mathrm{m}} )^2 \, \tau(t)}{a^3(\tau(t))} 
 \times \Delta_0^{-2}\, .
 \label{4pi}
 \end{equation}
 \noindent
Furthermore, by using $H^2_0(\tau_0)= 4/(1- \Omega_{\mathrm{m}})^2$ 
according to the expression for $H(\tau)$ and with the analytical solution 
$a(\tau) = (\frac{\tau^3- \Omega_{\mathrm{m}}}{1-\Omega_{\mathrm{m}}})^{2/3}$ 
obtained by \citet{Jesus17}, one finally has
 \begin{equation} 
 4 \pi G \varrho_{\mathrm{m}}(t) = \frac{6\, \Omega_{\mathrm{m}} \tau(t)}{\left(\tau^3(t)- \Omega_{\mathrm{m}}\right)^2}
 (1- \Omega^{\frac{1}{3}}_{\mathrm{m}} )^2 \times \Delta_0^{-2}\, ,
 \label{fin}
 \end{equation}
 \noindent
as a consistent expression of the density term in the Newtonian framework with the cosmic time as an independent variable. 
One can see that the references to the cosmic time entering in Equation (\ref{tt}) occur
through the expression of the Hubble parameter given by Equation (\ref{hh}). 

\section{Numerical results for the evolution of the density fluctuations} \label{num}

Let's examine the growth of the density fluctuations from the initial conditions in the CMB to the present time 
for several choices of parameters and initial conditions.



\subsection{Numerical values of the coefficients}

 The exact value of $\Omega_{\mathrm{m}}$ in the scale-invariant framework is unknown for now. 
 Therefore, one shall explore a large range of density parameters, say from $\Omega_{\mathrm{m}}=0.02$ to 0.50. Let us first
 examine the numerical values of parameters $H$, $\kappa$, $A$ and $B$ in the scale-invariant context for a chosen, rather conservative value of
 $\Omega_{\mathrm{m}}=0.10$, a value smaller than the one currently adopted
 since there is likely no need of dark matter within the scale-invariant context \citep{Maeder17c}. 
 Fig. \ref{Hk} shows the values of the Hubble constant 
 $H(t)$ and metrical function $\kappa(t)$ as functions of $\log (1+z)$ in this example.
 One can see that the value of $\kappa(t)$ is always rather close to $H(t)$, being 
 higher by about 0.176 dex for redshifts $z$ larger than about 100. 
 The separation between the two curves diminishes down to zero at
 $z+1 =1.978$, where the two curves cross each other and then $H(t)$ becomes slightly larger than $\kappa(t)$. 
Since $H$ and $\kappa$ are co-scalars of the same rank, one can use the analytical expressions in \citet{Jesus17} 
to obtain the value of $\tau$ when the crossing occurs, $\tau_{\times}=(1+\sqrt{5})/2\,\Omega_{\mathrm{m}}^{1/3}$ in the $\tau$-scale, 
and then the reciprocal of the scale factor $a(\tau)$ at that time gives $z+1=1/a(\tau)$.
This crossing point and its relevant values can be used to assess the accuracy of any numerical solutions.

 The trends of $H$ and $\kappa(t)$, illustrated in  Fig. \ref{Hk} for the same model as above, 
  determine the trends for the parameters $A$ and $B$ (Fig. \ref{AB}).
 The solutions of the equation (\ref{D4}) depend on the discriminant $\Delta$ of the equation. From (\ref{D4}), 
 one has $\Delta = \sqrt{A^2+4 B}$ and as illustrated in the example of Fig. \ref{AB}, the sum $A^2+4B$ is 
 always positive, which means that the solutions of the differential equation are real. 
 Let's concentrate on the growing solution, as is usually the case in the study of fluctuations.
 Parameter $A=2 H - 3 \kappa$ is  negative in the range of redshifts considered
 since $2 \kappa$ is always bigger than $H$. 
 While in the standard EdS model, the parameter $A$ is positive and acts as a damping effect,
 now the opposite is happening: $A$ is negative and favors the growth of the density fluctuations. In the  example of Fig. \ref{AB},
 parameter $ B=4 \pi G \varrho_{b} + 4 \kappa (H-\kappa) $ is negative for values above $z+1=3.16$, 
 the second term being larger than the first one at the low density considered. However, the constructive effect 
 of parameter $A$ dominates with respect to $B$ for the growth of the density fluctuations.

 \subsection{The initial conditions} \label{ini}
  
\begin{figure*}[h]
\centering
\includegraphics[width=.70\textwidth]{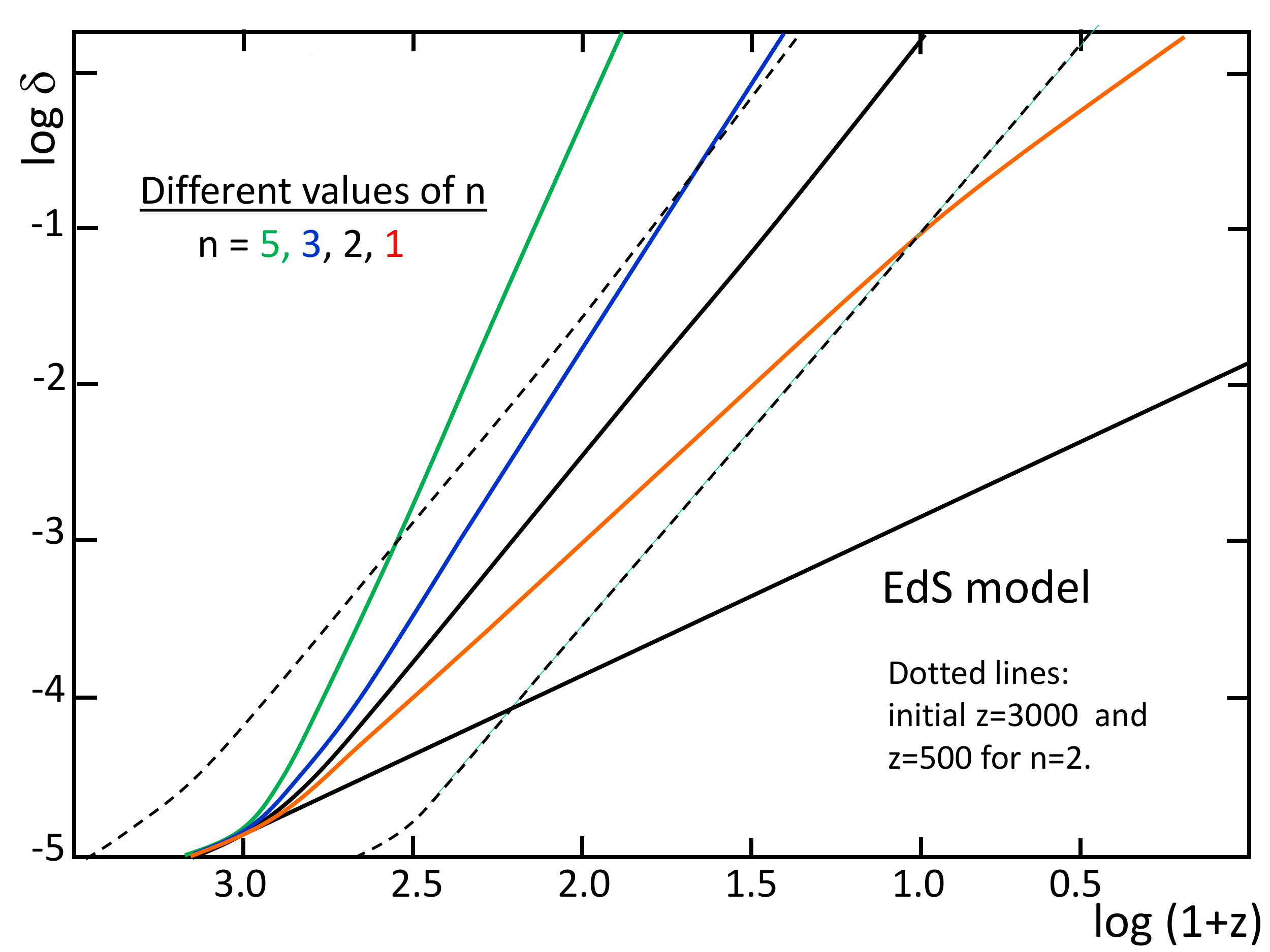}
\caption{The growth of density fluctuations for different values of parameter $n$ 
(the gradient of the density distribution in the nascent cluster), 
for an initial value $\delta= 10^{-5}$ at $z=1376$ and
$\Omega_{\mathrm{m}}=0.10$. The initial slopes are those of the EdS models.
The two light broken curves show models with initial $(z+1)=3000$ and 500,
with same $\Omega_{\mathrm{m}}=0.10$ and $n=2$. 
These dashed lines are to be compared to the black continuous line of the $n=2$ model. 
All the three lines for $n=2$ are very similar and nearly parallel.}
\label{variousn} 
\end{figure*}

 The amplitudes of the CMB fluctuations are about $\delta = 10^{-5}$ at the epoch of recombination
 that occurs at a redshift of the order of $z =10^{3}$. A value $\delta = 1.1 \cdot 10^{-5}$ is estimated at low angular separations, 
 while larger values (up to about $8 \cdot 10^{-5}$) are found at larger separations \citep{Hu02,Planck14,Clements17}.
 The estimate of the recombination epoch depends on the model and on the treatment of the physics of the recombination processes, 
 for example, the recombination is estimated to occur at a redshift $z= 1376$ by \citet{Durrer08}. In the scale-invariant model, 
 due to the difference of the scale-factors between the $\Lambda$CDM and the scale-invariant model \citep{Maeder17b}, it could be earlier, i.e around $z=1680$. 
 Evidently, the earlier is the start of the growing process, the larger is the final growth of the density perturbations. 
 Here, in order to be rather conservative let's keep the start of the growth at an initial redshift $z_{\mathrm{in}}=1376$ 
 with an initial value of $\delta= 10^{-5}$. However, the overall results are the same if one considers $z_{\mathrm{in}}=1000$ instead.
 Even, tests with $z_{\mathrm{in}}=3000$ and $z_{\mathrm{in}}=500$  are shown in Fig. \ref{variousn} to demonstrate the stability of the results.
 
To integrate a second-degree differential equation (\ref{D4}), one also needs the value of the derivative 
$\dot{\delta}$ at the starting point. For this purpose, we consider two different approaches to estimate the derivative 
$\dot{\delta}$ at the initial point. During the radiative era, the density fluctuations with a mass larger than the 
Silk mass survive without growing or declining. Thus, let us first assume that one has $\ddot{\delta}=0$ initially.
Thus, Equation (\ref{D4}) becomes
 \begin{equation}
 A \, \dot{\delta} \, \approx\, B \, \delta \, , \quad \mathrm{giving} \quad \frac{\delta_2}{\delta_1} \approx e^{\frac{B}{A} \Delta t} \,,
 \end{equation}
 \noindent
 where $\Delta t$ the initial time step $\Delta t= t_2 - t_1$ for the corresponding values of the perturbations
 $\delta_2$ and $\delta_1$. Let us consider a numerical example with the model for $\Omega_ {\mathrm{m}}=0.1$ and 
 parameter $n=2$.
 At time $t_1$ corresponding to $z+1= 1377$, we have $\delta_1=10^{-5}$, $A=-3.2773 \cdot 10^4$ and
 $B=-2.5779\cdot 10^8$. For a time step $\Delta t= 5.598 \cdot 10^{-7}$ in the cosmic timescale,
 we have $\frac{B}{A} \Delta t= 4.404\cdot 10^{-3}$
 and thus $\delta_2= 1.0044 \, \delta_1$ at redshift $z_2+1=1366.986$.
 Thus, for a ratio of the redshifts $\frac{z_1+1}{z_2+1} = 1.00733$, we have a corresponding ratio of the perturbations
 $\frac{\delta_2}{\delta_1}=1.00440$, which implies an initial slope $s$ of the relation $\log\delta$ vs. $\log (z+1)$
 equal to $s=0.60$, a bit lower than the slope $s=1$ of the classical EdS model for the growth of perturbations.
 
 As a second estimate, let us consider Equation (\ref{dn}) which says that the growth 
 of $\delta$ results from the motion gradient and from the cosmological 
 effect of scale invariance. Let us assume that the motion is initially negligible at the very beginning of the matter dominated era
 and that the growth is only due to the cosmological effect of scale invariance:
 \begin{equation}
 \dot{\delta} \, \approx \, n \, \kappa(t) \, \delta\, , \quad \mathrm{giving} \quad \frac{\delta_2}{\delta_1} 
 \approx e^{ n \, \kappa \Delta t} \, .
 \end{equation}
 \noindent
 Let us make an estimate for the same case as above. The cosmic time $t$ corresponding
 to $z+1= 1377$ is $ t=5.0853 \cdot 10^{-5}$ (with $t_0=1$). For the same $\Delta t$, we have
 $n \kappa(t) \Delta t= 2.202 \cdot 10^{-2}$ and $e^{ n \, \kappa \Delta t} = 1.0223 $, thus we have at time $t_2$
 a perturbation $\delta_2= 1.0223 \cdot 10^{-5}$. This ratio has to be compared to 
 $\frac{z_1+1}{z_2+1} = 1.00733$ and this gives a slope $s=3.03$, about three times larger than the 
 classical EdS slope.
 
 Thus, in what follows we shall adopt for the initial value of $\dot \delta$ the value corresponding to 
 $s=1$ of the EdS model. We will see that the true value is likely closer to the second estimate rather 
 than to the first one. However, this rather conservative
 choice $s=1$ allows us to not overestimate the initial growth of the density fluctuations. 
  Also,  relatively different choices of  the adopted  slope at the initial $z$ lead  to similar results  
  when $\delta$ becomes larger than about $10^{-4}$ (see also Sect. \ref{lin}).  
Finally, we note that the oscillations occurring in the radiative era also provide significant values of $\dot{\delta}$ 
\citep{Coles02}, which may also contribute to ``the launch'' of the growth of the density fluctuations.
 
\begin{figure*}[h]
\centering
\includegraphics[width=.70\textwidth]{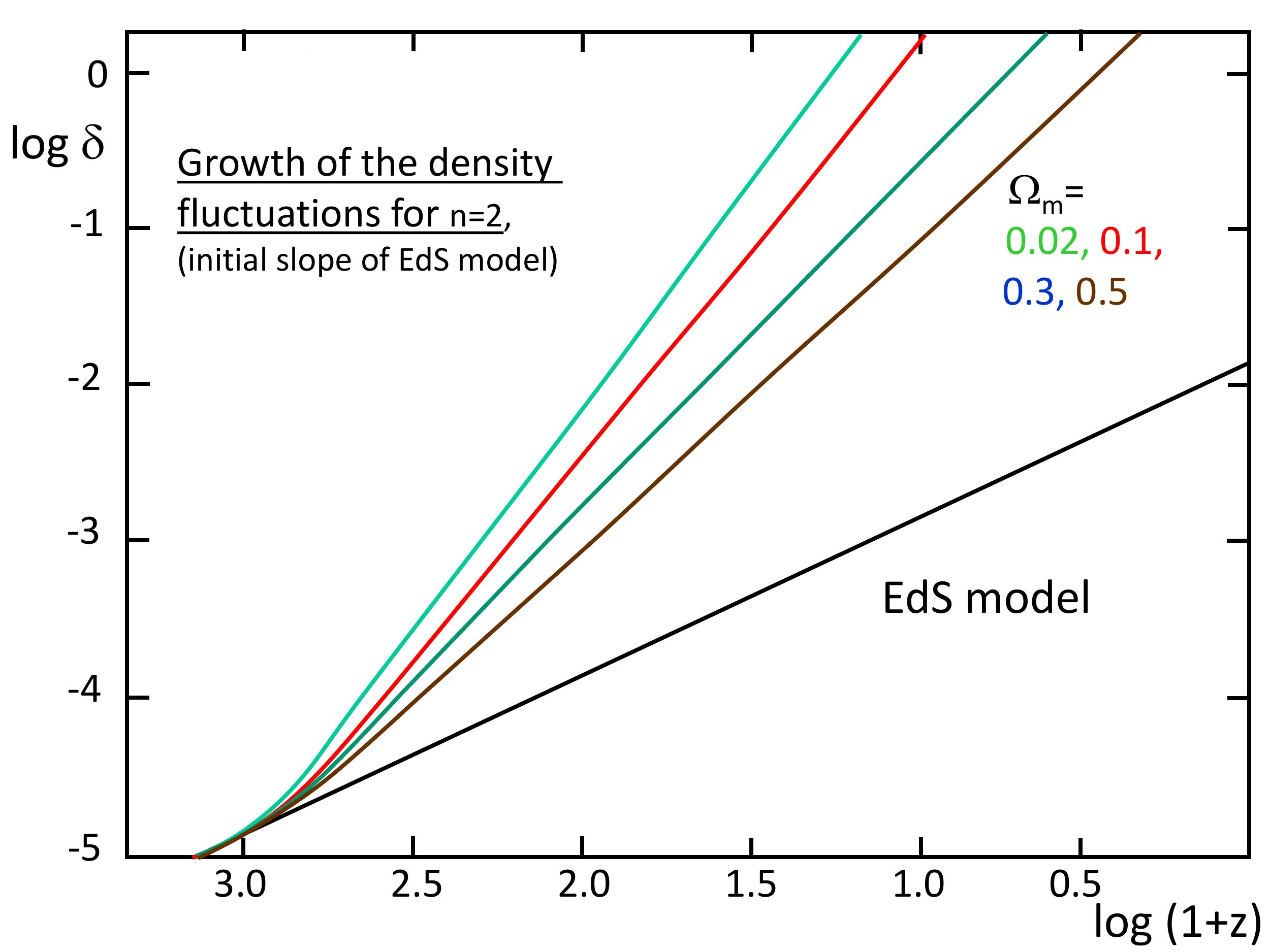}
\caption{The black curve is the classical growth of the density fluctuations in the Einstein-de Sitter model. 
The other four curves illustrate the growth of $\delta$ for the density profile with $n=2$ in the scale-invariant theory. 
There are four different values of the density parameter $\Omega_{\mathrm{m}}$. 
An initial value $\delta= 10^{-5}$ at $z=1376$ has been taken for all models, the initial derivative 
$\dot{\delta}$ is taken equal to that of the EdS model. After a short evolution with a slope close 
to the initial one, all solutions indicate a much faster growth of the density fluctuations,
reaching the non-linear regime between about $z+1= 2.7$ and $z=18$.}
\label{deredsn2} 
\end{figure*}

\subsection{The almost linear growth of the density fluctuations} \label{lin}

As discussed above,  we are choosing the low slope present in the standard EdS model as the initial value of 
$\dot{\delta} (z_{\mathrm{in}})$.
 The EdS model behaves like $\delta \propto t^{2/3} \propto 1/(z+1)$ and the corresponding 
 slope ($s=1$ in the $\log \delta$ vs. $\log (z+1)$) is illustrated in
 Fig. \ref{deredsn2} in the case of $n=2$. Scale-invariant solutions for the growth of the density fluctuations have been calculated
 for $\Omega_{\mathrm{m}}=0.02, 0.10, 0.30$ and 0.50, in order to encompass a large range of possibilities.
 Interestingly enough, after having followed for a short time the initial slope of the EdS model,
 the resulting scale-invariant solutions show a much faster growth of 
 the perturbation $\delta$. Some small deviations from the linearity appear below $\delta \approx 10^{-4}$, 
 where the slope is slightly lower since it is still influenced by the initial value adopted.
 Above a value of about $\delta \approx 10^{-4}$, these steep growths follow almost
 linear relations in the $\log \delta$ vs. $\log (z+1)$ plot with slopes $s$ that we will call the {\it convergence slopes}.
 In a few cases, there are some deviations from linearity in the relations $\log \delta$ vs. $\log (z+1)$, 
  in particular  for the case $n=1$ at the lowest
 redshifts as illustrated in Fig. \ref{variousn}. 
 This indicates that in general the growth of the density fluctuations is very close to a power law dependence like 
 $ \delta \sim \left(\frac{1}{(z+1)}\right)^s$. The convergence slopes of the linear relations in the log scales
 are $s=1.95, 2.22, 2.62$ and 2.90 
 for $\Omega_{\mathrm{m}}= 0.50, 0.30, 0.10$ and 0.02 respectively, in the case $n=2$. 
 For the same values of $\Omega_{\mathrm{m}}$ and $n=3$, these slopes are 2.25,
 2.70, 3.35 and 3.91. 
 
 We notice that lower matter density experiences a faster growth of the density perturbations. This results from the changes 
 of $H$ and $\kappa$ with $\Omega_{\mathrm{m}}$ intervening in 
 the leading parameters $A$ and $B$. In the early evolution of the perturbation, the values of $H$ and $\kappa$ are larger
 (factor about 5) in model with $\Omega_{\mathrm{m}}=0.50 $ than with 0.02, 
 while the ratio between $H$ and $\kappa$ has small variations. 
 The square of this factor (of $\sim 5$) intervenes in parameter $B$, while the effect is linear for $A$. Thus, the effect of a larger 
 $\Omega_{\mathrm{m}}$ is more prominent in the term $B$ that is contributing negatively to the perturbation than 
 in the term $A$ which is contributing positively. This accounts for the slower growth of the perturbations in higher density models, despite
 the higher Newtonian attraction (which varies linearly with $\Omega_{\mathrm{m}}$).
 
Another way to see and understand the contra-intuitive behavior of the growth of the perturbations with respect to $\Omega_{\mathrm{m}}$
is to notice that $\kappa>H$ for most of the early epoch as shown in Fig. \ref{Hk}. 
Thus for $n\geq 1$, the term $A<0$ and therefore has an anti-dumping effect until the $\kappa\sim H$ epoch. 
The $B$ term, on the other hand, is negative at the very early times $t\gtrapprox 0$ due to the $\kappa^2$ term. 
This $\kappa^2=1/t^2$ term is the same for any $\Omega_{\mathrm{m}}$. However, $H$ is bigger for smaller  
$\Omega_{\mathrm{m}}$ as seen in Table 1 \citep{Maeder17a}.
Thefore, the smaller the $\Omega_{\mathrm{m}}$, the smaller the negative term $B$, 
since $H$ appears as a positive term in this negative $B$ expression.
Thus, the growth is controlled by the $A$ term while the suppressing effects of the negative $B$ term 
are smaller for smaller $\Omega_{\mathrm{m}}$ values.
 
 We also find that largely  different initial slopes result in similar convergence slopes.
 For example, a model for $\Omega_{\mathrm{m}}=0.10$ with $n=2$ with an initial  slope 5 times smaller 
 than the corresponding convergence slope
 shows deviations from the convergence slope smaller than 1\% in the range of $\log \delta $ from -3 to -1.
 
 With the initial slope of the EdS model, a density fluctuations $\delta =1$ is reached for redshifts between $(z+1)= 2.7$ and 18 for $n=2$,
 depending on the $\Omega_{\mathrm{m}}$ values (Fig. \ref{deredsn2}). 
 During the subsequent evolution towards lower redshifts, the growth of the density fluctuations become non-linear and may lead to 
 the present day inhomogeneous Universe.
 
\begin{figure*}[h]
\centering
\includegraphics[width=.70\textwidth]{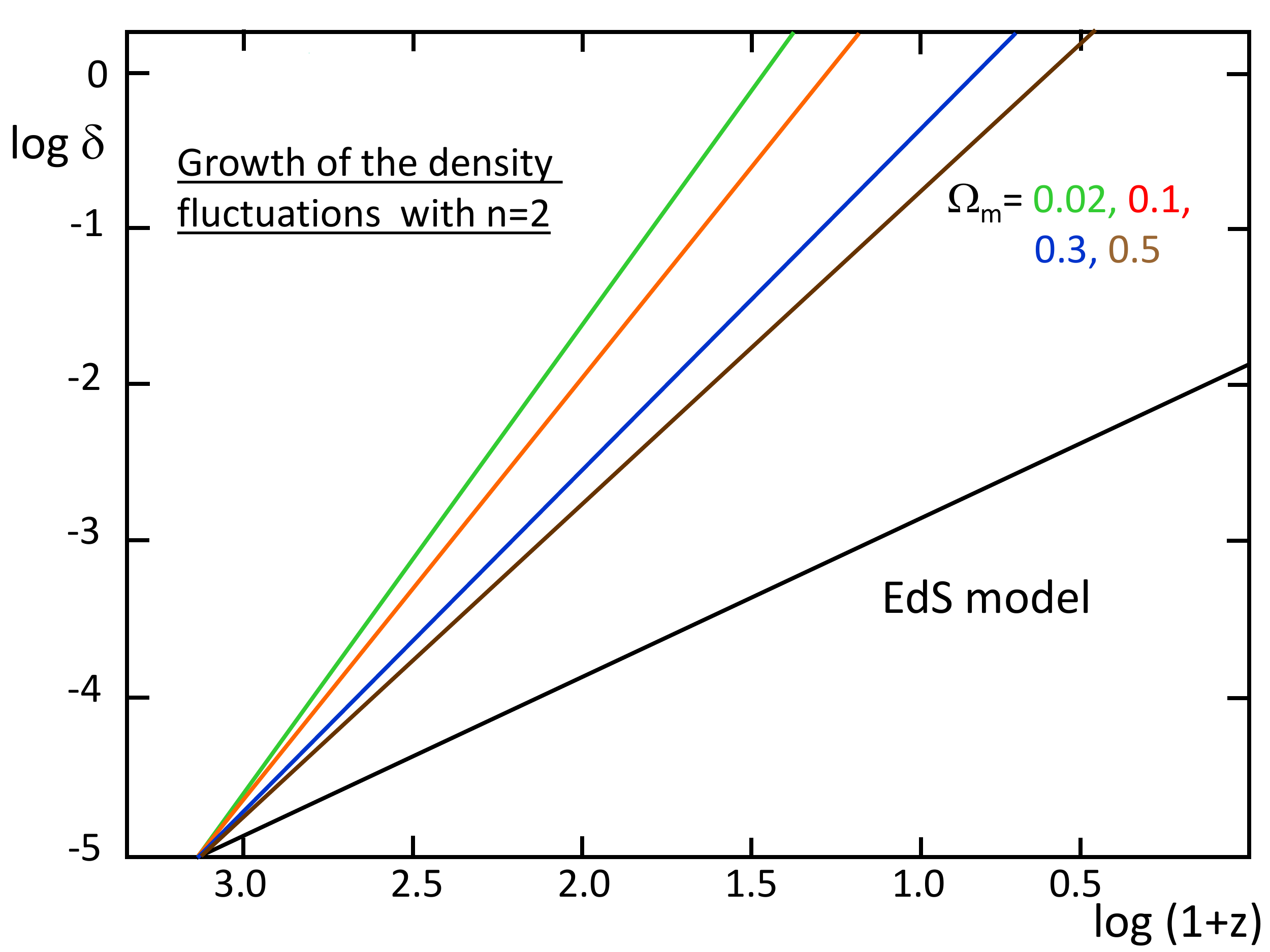}
\caption{The growth of density fluctuations for different values of parameter $\Omega_{\mathrm{m}}$ for $n=2$.
The initial slopes of the density fluctuations have been taken from the convergence slopes of the linear domain 
(Fig. \ref{deredsn2}) of the models with the corresponding $\Omega_{\mathrm{m}}$. 
The values $\delta=1$ are reached in the interval of redshifts $(z+1)=4.0$ to 29.3.}
\label{contn2} 
\end{figure*} 

\subsection{Discussion of the tests and results} \label{res}

 We now discuss the dependence of the evolution of the perturbations on the various parameters 
 characterizing the models: the matter density in the Universe $\Omega_{\mathrm{m}}$, the slope $n$ of the
 distribution of density in the forming cluster, the initial values
 of $\delta$ at $z$ equal to about 1000 and the value of this initial $z$ redshift.
  
 The effects of different slopes $n$ of the density distribution in the infalling body are illustrated in Fig. \ref{variousn}, which show
 results for $n=1, 2 , 3$ and 5 with still the initial EdS slopes.
 We see that steeper density gradients produce a faster growth of the density fluctuations and thus an earlier entering into the 
 nonlinear domain, which marks the time of galaxy formation. For $\Omega_{\mathrm{m}}=0.10$, 
 from a value $n=1$ to $n=5$, the redshift when $\delta=1$ is reached changes from
 $(z+1)=2.2$ to about 88 (Fig. \ref{variousn}). 
 Whether due to local effects the representative value of $n$ may vary in forming clusters of galaxies 
 is an open question, beyond the scope of the present work. But, it opens the possibility that, in some cases,
 galaxies could form very early in the Universe!
 
 The initial value of $\delta$ does not influence the relative growth
 over a given interval of redshift. For example, a fluctuation $\delta= 10^{-4}$ at $z=1000$ will evolve all the way being
 always a factor of 10 larger than those created by an initial  fluctuation $\delta= 10^{-5}$, at least as long as in the linear regime.
 Thus, the relative growth of the linear fluctuations predicted by the Equation (\ref{D4}) is also scale-invariant since $\delta$ is an in-scalar. 
 The value of the initial slope has been discussed in 
 Sec. \ref{ini}. 
 
 Finally, we also point out that the value of the initial $z$ has little influence on the convergence slope. This is nicely illustrated 
 by the two thin black broken lines in Fig. \ref{variousn}. For the purpose of the demonstration,
 one model is starting with a fluctuation $\delta=10^{-5}$ at $(z+1)=3000$
 and another one at $(z+1)=500$. This is to be compared to the standard case with $(z+1)=1377$. 
 We notice an almost parallel evolution of the growing perturbations in the three models.

Consistently with previous remarks, the value for the convergence slopes are almost independent of the initial redshift, 
whatever is its particular value. Fig. \ref{contn2} illustrates the results for models with $n=2$ and 
different $\Omega_{\mathrm{m}}$ values starting at $z=1376$. 
For this case, discussed in Sec. \ref{lin}, the convergence slope determined for the models of various density parameters 
(Fig. \ref{deredsn2}) has been used to fix the initial derivative or slope of the corresponding model for the growth of density fluctuations.
Fig. \ref{contn2} shows that the relations are linear from the initial redshift to the present epoch. Globally, the results confirm those 
of Fig. \ref{deredsn2}, with the differences that due to the initial faster growth, the amplitudes $\delta=1$
are reached a bit earlier. Values $\delta=1$ are reached for values between $(z+1)= 4.0$ and 29.3 for the different 
$\Omega_{\mathrm{m}}$ (Fig. \ref{contn2}). 
Evidently, the simple extrapolation of the scale-invariant models up to the present time would lead to enormous values of the density
fluctuations $\delta$. However, this does not apply, because in the non-linear regime many other effects intervene.
On one side, the growth may become initially faster, but also at some stage pressure effects may intervene limiting the growth
of density fluctuations.
 
The general result is that the scale-invariant dynamics permits the growth of the density fluctuations
from $\delta \approx 10^{-5}$ on the last scattering surface to the significant inhomogeneities present in 
today's Universe. In realistic cases, the growth of the density fluctuations is very fast, the regime where 
$\delta >1$ is typically entered at redshifts of a few tens depending on the average background density.

\section{Conclusions} \label{concl}

In this work, the continuity equation, the equations of Euler and Poisson have been expressed in the scale-invariant context.
These equations were then utilized to derive Equation (\ref{D2}) that governs the evolution of the density perturbations 
in a matter-dominated universe. Numerical solutions are obtained for an initial amplitude
$\delta=10^{-5}$ at redshift $z \approx 10^3$ and for various density parameters $\Omega_{\mathrm{m}}$. 
In the standard model of a matter-dominated Universe, the fluctuations of density $\delta$ grow like $(\frac{1}{1+z})^s$ 
with $s=1$. In the scale-invariant case, the growth of density fluctuations is much faster. 
The values of $s$ typically range from about 2.2 to 2.9 for $\Omega_{\mathrm{m}}$ 
between 0.30 and 0.02. This enables the density fluctuations to enter the regime with 
$\delta > 1$ long before the present time, typically at redshifts of an order of 10 or more.
 
Thus, to form the galaxies and clusters of galaxies, the baryons do not need 
to settle down in a potential well previously installed by an unknown form of dark matter component, since
the growth of the density fluctuations in the scale-invariant context is significant and fast enough.
This result is in agreement with previous works \citep{Maeder17c, Maeder18} showing that dark matter is not needed
to account for the high velocities observed in clusters of galaxies, the high rotation
velocities of stars at the edge of spiral galaxies, or the growth of the ``vertical" velocity dispersion 
of stars as correlated to their age within our Galaxy.
 
 \vspace*{1cm} 
Acknowledgments: 
A.M. expresses his deep gratitude to his wife and to D. Gachet for their continuous support.
V.G. is extremely grateful to his wife and daughters for their understanding and family support 
during the various stages of the research presented.

This research did not receive any specific grant from funding agencies in the public, commercial, or not-for-profit sectors.


\end{document}